\documentstyle[onecolumn,epsfig]{mn}
\newif\ifAMStwofonts

\newcommand{\etal}{{et~al.}}

\newcommand{\lsim}{\,\lower2truept\hbox{${<\atop\hbox{\raise4truept\hbox{$\sim$}}}$}\,}
\newcommand{\gsim}{\,\lower2truept\hbox{${>\atop\hbox{\raise4truept\hbox{$\sim$}}}$}\,}

\def\lsim{\,\lower2truept\hbox{${< \atop\hbox{\raise4truept\hbox{$\sim$}}}$}\,}
\def\gsim{\,\lower2truept\hbox{${> \atop\hbox{\raise4truept\hbox{$\sim$}}}$}\,}
\def\Ohat{{\widehat \Omega}}

\def\deg{\ifmmode^\circ \else$^\circ $\fi}    
\def\arcs{\ifmmode {'' }\else $'' $\fi}     
\def\arcm{\ifmmode {' }\else $' $\fi}     
\def\buildrel#1\over#2{\mathrel{\mathop{\null#2}\limits^{#1}}}
\def\mper{\ifmmode \buildrel m\over . \else $\buildrel m\over .$\fi}
\def\hper{\ifmmode \rlap.^{h}\else $\rlap{.}^h$\fi}
\def\sper{\ifmmode \rlap.^{s}\else $\rlap{.}^s$\fi}
\def\arcsper{\ifmmode \rlap.{' }\else $\rlap{.}' $\fi}
\def\arcmper{\ifmmode \rlap.{'' }\else $\rlap{.}'' $\fi}

\def\mincir{\ \raise -2.truept\hbox{\rlap{\hbox{$\sim$}}\raise5.truept	
\hbox{$<$}\ }}								%
\def\magcir{\ \raise -2.truept\hbox{\rlap{\hbox{$\sim$}}\raise5.truept	%
\hbox{$>$}\ }}								%

\def\etal   {et~al.\,}

\title[Early and late CMB spectral distortions and millimetric foreground]
{A joint study of early and late spectral distortions of the 
cosmic microwave background and of the millimetric foreground}

\author[R. Salvaterra \& C. Burigana]
{R. Salvaterra$^1$ \& 
C. Burigana$^2$\\
$^1$SISSA/ISAS, Astrophysics Sector, Via Beirut, 4, I-34014 Trieste, Italy \\
$^2$IASF/CNR, Istituto di Astrofisica Spaziale e Fisica Cosmica,
Sezione di Bologna,\\ 
Consiglio Nazionale delle Ricerche, 
Via Gobetti 101, I-40129 Bologna, Italy}

\date{Submitted to MNRAS, 14 March 2002; in this revised version, 17 June 2002;
accepted, 19 June 2002.}

\pagerange{\pageref{firstpage}--\pageref{lastpage}}
\pubyear{2002}

\begin{document}

\maketitle

\label{firstpage}
\footnotetext{E-mail: salvater@sissa.it ; burigana@tesre.bo.cnr.it}
\begin{abstract}
In this work we have compared the absolute temperature data 
of the CMB spectrum with models of CMB spectra distorted by a single or two 
heating processes at different cosmic times. 
The constraints on the fractional energy injected
in the radiation field, $\Delta \epsilon/\epsilon_i$,
are mainly provided by the precise measures of the FIRAS instrument aboard 
the COBE satellite, while long wavelength measures are crucial 
to set constraints on free-free distortions.
We find that the baryon density
does not influence the limits on $\Delta \epsilon/\epsilon_i$
derived from current data for cosmic epochs
corresponding to the same dimensionless time $y_h$ of dissipation epoch, 
although the redshift corresponding to the same $y_h$ decreases with the baryon density.
Under the hypothesis that two heating processes 
have occurred at different epochs, 
the former at any $y_h$ in the range $5 \geq y_h \geq 0.01$ 
(this joint analysis is meaningful for $y_h \gsim 0.1$)
and the latter at $y_h \ll 1$, 
the limits on $\Delta\epsilon/\epsilon_i$ are relaxed by a factor $\sim 2$ 
both for the earlier and the later process with respect to the case in which 
a single energy injection in the thermal history of the universe is considered.
In general, the constraints on  $\Delta\epsilon/\epsilon_i$
are weaker for early processes
($5 \gsim y_h \gsim 1$) than for relatively late processes
($y_h \lsim 0.1$), because of the sub-centimetric wavelength coverage
of FIRAS data, relatively more sensitive to Comptonization
than to Bose-Einstein like distortions.

While from a widely conservative point of view 
the FIRAS calibration as revised by Battistelli et al. 2000
only implies a significant relaxation 
of the constraints on the Planckian shape of the CMB spectrum,
the favourite calibrator emissivity law proposed by the authors,
quite different from a constant emissivity,
implies significant deviations from a Planckian spectrum.
An astrophysical explanation of this, although intriguing, 
seems difficult.
We find that an interpretation in terms of CMB spectral distortions
should require a proper balance between the energy exchanges at two 
very different cosmic times
or a delicate fine tuning of the parameters characterizing a dissipation 
process at intermediate 
epochs, while an interpretation in terms of a relevant millimetric 
foreground, produced by cold dust, should imply
a too large involved mass and/or an
increase of the fluctuations at sub-degree angular scales.
Future precise measurements at longer wavelengths
as well as current and future CMB anisotropy space missions
will provide independent, direct or indirect, cross checks.

This work is related to {\sc Planck}-LFI activities. 
\end{abstract}


\section{Introduction}

As widely discussed in many papers, the spectrum of the cosmic microwave 
background (CMB) carries unique informations on physical processes occurring
at early cosmic epochs
(see, e.g., Danese \& Burigana 1993 and references therein).
The comparison between theoretical models of CMB spectral distortions 
and CMB absolute temperature measures can constrain the 
physical parameters of the considered dissipation processes.
We improve here the previous analyses of the implications of the CMB spectrum data
by jointly considering distortions generated in a wide range of 
early or intermediate cosmic epochs and at late cosmic epochs. 
We consider also the implications of a recent analysis of the COBE/FIRAS calibration
by Battistelli et al. 2000,
which suggests a decrement of the absolute temperature 
with the wavelength.

In section~2 we briefly summarize the general properties of the
CMB spectral distortions, the main physical informations that can be derived from
the comparison with the observations, and the relationship between 
the detection of possible CMB spectral distortions and the evaluation
of the level of astrophysical sub-millimetric and millimetric foregrounds.
The data sets used in this study are presented in section~3.
In section~4 we present the constraints on CMB spectral distortions
based on ground and balloon experiments and on the ``standard'' COBE/FIRAS
data calibrated according to the COBE/FIRAS team (Mather et al. 1999 and references therein)
for the case of a single and of a double
energy injection in the thermal history of the universe.
The extrapolation at very high redshifts and the impact of free-free distortions
are considered in sections~5 and 6. 
In section~7 we analyse the implications of the calibration by 
Battistelli et al. 2000, by briefly reporting the main properties 
of their revised monopole spectral shape, 
and by jointly evaluating its impact on
the CMB spectral distortion constraints and for the millimetric foreground.
Finally, we draw our main conclusions in section~8.

\section{Theoretical framework and data exploitation}

The CMB spectrum emerges from the thermalization redshift, 
$z_{therm} \sim 10^6 \div 10^7$, 
with a shape very close to a Planckian one, 
owing to the strict coupling between radiation and matter through
Compton scattering and photon production/absorption processes, 
radiative Compton and Bremsstrahlung,
which were extremely efficient at early times 
and able to re-establish a blackbody (BB) spectrum 
from a perturbed one
on timescales much shorter than the expansion time (see, e.g., 
Danese \& De~Zotti 1977).
The value of $z_{therm}$ (Burigana et~al. 1991a)
depends on the baryon density (in units of the critical density),
$\Omega_b$, 
and the Hubble constant, $H_0$, through the product 
$\Ohat_b =\Omega_b (H_{0}/50)^2$ ($H_0$ expressed in Km/s/Mpc). 

On the other hand, physical processes occurring at redshifts $z < z_{therm}$ 
may lead imprints on the CMB spectrum.

\subsection{General properties of the CMB spectral distortions}

The timescale for the achievement of
kinetic equilibrium between radiation and matter
(i.e. the relaxation time for the photon spectrum), $t_C$, is
\begin{equation}
t_C=t_{\gamma e} {m c^{2}\over {kT_e}} \simeq 4.5 \times 10^{28} 
\left( T_{0}/2.7\, K \right)^{-1} \phi ^{-1} \Ohat_b^{-1}
\left(1+z \right)^{-4} \sec \, ,
\end{equation}
where $t_{\gamma e}= 1/(n_e \sigma _T c)$ is the photon--electron collision
time, $\phi = (T_e/T_r)$, $T_e$ being the electron temperature and
$T_r=T_{0}(1+z)$;
$kT_e/mc^2$ is the mean fractional change of photon energy in a scattering
of cool photons off hot electrons, i.e. $T_e \gg T_r$;
$T_0$ is the present radiation temperature related
to the present radiation energy density by $\epsilon _{r0}=aT_0^4$;
a primordial helium abundance of 25\% by mass is here assumed.

It is useful to introduce the dimensionless time variable $y_e(z)$ defined by
\begin{equation}
y_e(z) = \int^{t_0}_{t} {dt \over t_C}
=\int^{1+z}_{1} {d(1+z) \over 1+z} {t_{exp}\over t_C} \, ,
\end{equation}
where $t_0$ is the present time and
$t_{exp}$ is the expansion time given
by
%
\begin{equation}
t_{exp} \simeq   6.3\times 10^{19} \left({T_0 \over 2.7\, K}\right)^{-2}
(1+z)^{-3/2} \left[\kappa (1+z) + (1+z_{eq})
-\left({\Omega_ {nr} -1 \over \Omega_ {nr}}\right)
\left({1+z_{eq} \over 1+z}\right) \right]^{-1/2}
\sec \, ,
\end{equation}
$z_{eq} = 1.0\times 10^4 (T_{0}/2.7\, K)^{-4}\Ohat _{nr}$
being the redshift of
equal non relativistic matter and photon energy densities
($\Omega _{nr}$ is the density of non relativistic matter in units of critical
density); $\kappa = 1 + N_\nu (7/8)
(4/11)^{4/3}$, $N_\nu$ being the number of relativistic, 2--component,
neutrino species (for 3 species of massless neutrinos, $\kappa \simeq 1.68$),
takes into account the
contribution of relativistic neutrinos to the dynamics of the
universe\footnote{Strictly speaking the present ratio of neutrino to
photon energy densities, and hence the value of $\kappa$, is itself a
function of the amount of energy dissipated. The effect, however,
is never very important and is negligible
for very small distortions.}.

Burigana et al. 1991b have reported on
numerical solutions of the Kompaneets equation (Kompaneets 1956)
for a wide range of values of the relevant parameters.

The analysis of the constraints on the thermal
history of the universe set by the
high accuracy measurements that have been recently accumulated requires
the use of manageable formulae describing spectral distortions
for a wide range of the relevant parameters.

Under the assumptions of $i)$ small distortions,
$ii)$ dissipative processes with negligible
photon production, $iii)$ heating close to be instantaneous,
a good approximation if the timescale for energy dissipation is much
smaller than the expansion timescale, $iv)$ distorted radiation spectrum
initially represented by a superposition of blackbodies,
as is the case for a broad variety of situations of cosmological interest, 
Burigana et al. 1995
found accurate analytical representations of the numerical solutions 
for the photon occupation number
$\eta$ computed by Burigana et al. 1991b. 
The CMB distorted spectra depend on at least
three main parameters: the fractional amount of energy exchanged between
matter and radiation, $\Delta\epsilon / \epsilon_i$,
$\epsilon _i$ being the radiation energy density before the energy injection,
the redshift $z_h$ at which the heating occurs, and the
baryon density $\Ohat_b$.
The photon occupation number can be then expressed in the form
\begin{equation}
\eta = \eta (x; \Delta\epsilon / \epsilon_i, y_h, \Ohat_b) \, ,
\end{equation}
where $x$ is the dimensionless frequency $x = h\nu/kT_{0}$
($\nu$ being the present frequency),
and $y_h \equiv y_e(z_h)$ characterizes the epoch when the energy dissipation
occurred, $z_h$ being the corresponding redshift
(we will refer to $y_h \equiv y_e(z_h)$ computed assuming $\phi=1$, so that the epoch 
considered for the energy dissipation does not depend on the 
amount of released energy).

The form of these analytical approximations is in part suggested by 
the general properties of the Kompaneets equation and by its
well known asymptotic solutions.
For late distortions ($y_h \ll 1$) a superposition of
blackbodies is, to a very good approximation, a solution of the
Kompaneets equation, except at very low frequencies where photon emission
processes are important; when they dominate
the Kompaneets equation reduces to an ordinary differential
equation. 
The Comptonization distortion produced by hot gas at small $z$ is 
a typical example of superposition of blackbodies 
(Zeldovich \& Sunyaev 1969; Zeldovich et al. 1972).
At the other extreme (early distortions, $y_h \gsim 5$) the solution is well
described by a Bose-Einstein (BE) formula with a frequency dependent
chemical potential. For intermediate values of $y_h$,
$\eta$ has a shape somewhere between these two limiting
cases. The shape of the distorted spectra at long 
wavelengths is characterized by a spread minimum of the brightness temperature
for $y_h \lsim 0.5$, and by a minimum at a well defined wavelength for $y_h \gsim 0.5$.
The continuous behaviour of the distorted spectral shape with $y_h$ can be
in principle used also to search for constraints on the epoch of the energy exchange.

Of course, by combining the approximations describing the distorted spectrum
at early and intermediate epochs with the Comptonization distortion expression
describing late distortions, we are able to jointly treat two heating 
processes.

We remember that the adopted analytical description 
of the distorted spectrum holds also for continuous heating processes,
i.e. by relaxing the above assumption $iii)$,
provided that they occur at quite late
epochs ($y_h \lsim 0.05$) or at early epochs ($y_h \gsim 5$), because of the
properties of the asymptotic solutions.
In the former case, $\Delta\epsilon / \epsilon_i$ is the 
dissipation rate integrated over the relevant time interval;
in the latter case, the proper value of $\Delta\epsilon / \epsilon_i$ 
to be used in eq.~(4) is given by $\simeq \mu (y_h \simeq 5) /1.4$, where
$\mu (y_h \simeq 5)$ is the chemical potential at $y_h \simeq 5$
produced by the combined effect of the dissipation rate 
integrated over the time (which increases $\mu$) and of 
radiative Compton and Bremsstrahlung
(which decrease $\mu$).
In this case, eq.~(4) represents a good approximation of the distorted
spectrum  by relaxing also the above assumption $ii)$ provided that
the combined effect of energy injection and photon production
is properly included in the computation of the chemical 
potential $\simeq \mu (y_h \simeq 5)$ (see, e.g., Danese \& Burigana 1993). 
For continuous processes at intermediate epochs ($5 \gsim y_h \gsim 0.05$),
the above representation, with a proper ``effective'' value 
of $y_h$, can be used as a first order approximation of the
distorted spectrum.
Finally, for dissipation processes widely distributed in time,
from $y_h \gsim 5$ to $y_h \ll 1$ (as some of those discussed in section~7.6,
such as the damping of perturbations and the vacuum decay),
the detailed final spectrum shape will depend on the 
details of the considered process and will show
BE-like and/or Comptonization features more or less relevant 
according to the amount of energy dissipated at early and/or late epochs.

\subsection{Comparison between observations and models} 

We compare the measures of the CMB absolute temperature,
briefly summarized in section~3,
 with the above models
of distorted spectra for one or two heating processes
by using a standard $\chi^2$ analysis. 

We determine the limits on the amount of energy possibly injected in the cosmic
background at arbitrary primordial epochs corresponding to a redshift $z_h$ (or 
equivalently to $y_h$).
This topic has been discussed in several papers
(see, e.g., Burigana et al. 1991b, 
Nordberg \& Smoot 1998). We improve here the previous methods of analysis 
by investigating the possibility of properly combining 
FIRAS data with longer wavelength measurements and by refining the method 
of comparison with the theoretical models. We will consider the recent improvement in the 
calibration of the FIRAS data, that sets the CMB scale temperature to
$2.725\pm0.002$~K at 95\% CL (Mather et al. 1999). We consider 
the effect on the estimate of the amount of energy injected in the CMB 
at a given epoch introduced by the calibration uncertainty of FIRAS scale temperature 
when FIRAS data are treated jointly to longer wavelength measures.
Thus, we investigate the role of available ground and balloon 
data compared to the FIRAS measures. 

Then, we study the combined effect of two different heating processes
that may have distorted the CMB spectrum at different epochs.
This case has been also considered in the paper by Nordberg \& Smoot 1998, where 
the CMB absolute temperature data are compared with theoretical spectra 
distorted by a first heating process at $y_h=5$, 
a second one at $y_h\ll 1$ and by free-free emission, to derive limits on the
parameters that describe these processes. 
We extend their analysis by 
considering the full range of epochs for the early and intermediate energy injection
process, by taking advantage of the analytical representation 
of spectral distortions at intermediate redshifts (Burigana et~al. 1995).
Also in this case, 
the analysis is performed by taking into account the FIRAS calibration uncertainty.

In each case, we fit the CMB spectrum data 
for three different values, 0.01, 0.05 and 0.1,
of the baryon density  $\Ohat_b$.
In presence of an early distortion,
$\Ohat_b$ could be in principle estimated by CMB spectrum observations 
at long wavelengths, able to detect the wavelength of 
the minimum of the absolute temperature,
determined only by the well known physics of the radiation
processes in an expanding universe during the radiation dominated era.

We present our results on the above arguments in section~4.

Then, we extend in section~5 the limits on $\Delta\epsilon/\epsilon_i$ for energy injection
processes possibly occurred at $z_h>z_1$, being $z_1$ is the redshift corresponding
to $y_h = 5$, when the Compton 
scattering was able to restore the kinetic equilibrium between matter and
radiation on timescales much shorter than the expansion time
and the evolution on the CMB spectrum can be easily studied
by replacing the full Kompaneets equation with the differential
equations for the evolution of the electron temperature and the chemical potential.
This study can be performed by using the 
simple analytical expressions by Burigana et al. 1991b
instead of numerical solutions.
For simplicity, we restrict this analysis to the case $\Ohat_b=0.05$ 
and to the best-fit value of the FIRAS calibration. 

The relationship between the free-free distortion and the Comptonization
distortion produced by late dissipation processes 
depends on the details of the thermal history at late epochs
(Danese \& Burigana 1993, Burigana et al. 1995)
and can not simply represented by integral parameters. In addition,
free-free distortions are particularly important at very long
wavelengths, where the measurements 
have the largest error bars, at least for energy injection processes
which give positive distortion parameters;
for cooling processes, which generate negative distortion parameters,
the effect may be more relevant also at centimetric
wavelengths, but the connection between free-free and Comptonization
distortions becomes even more crucial.
Therefore, we firstly carry out the above analyses of early/intermediate 
and late distortions by neglecting free-free distortions, i.e. 
assuming a null free-free distortion parameter $y_B$.
This kind of distortion as well as its impact
on the constraints derived for the energy injected at different cosmic times
is considered in section~6.

For sake of completeness,
we finally observe that negative distortion parameters can be produced by
physical processes (e.g., cooling processes, radiative decays of massive particles)
that in general are described by a set of process parameters more complex than 
that considered here (epoch and energy exchange only) and produce 
spectral shapes different than those considered here. Therefore,
the constraints derived for negative values of 
$\Delta \epsilon/\epsilon_i$ and $y_B$ have to be considered 
only as indicative.

For compactness, we avoid to report in sections~4 and 6 
the fit results for $T_0$ which
is found to be only just different from the FIRAS calibration temperature scale,
according to the considered data set and fit parameters.
On the contrary, in section~7, where we analyse the implications of the 
FIRAS calibration as revised by Battistelli et~al. 2000, we report also 
the best fit values found for $T_0$ in terms of $\Delta T_0 = T_0-2.725$K.

\subsection{Sub-millimetric and millimetric foregrounds}

A crucial step for the analysis of the CMB spectral distortions 
is the subtraction of the astrophysical monopole from the total monopole
signal. At sub-millimetric wavelengths the integrated contribution from 
unresolved distant galaxies is expected to significantly overwhelm 
the difference between the intensities of a distorted spectrum and 
a blackbody spectrum with the same, or very close, $T_0$.
This greatly helps the extraction 
of the sub-millimetric foreground from the total monopole.
Three independent methods to extract the sub-millimetric extragalactic 
monopole from FIRAS data 
in the ``Low'' (LLSS, 30--660 GHz) and ``High'' (RHSS, 60--2880 GHz) frequency bands
(see COBE/FIRAS Explanatory Supplement 1995) 
carried out in the recent past by Puget et~al. 1996, 
Burigana \& Popa 1998 and Fixsen et~al. 1998 are in good agreement one each other;
in particular, the sub-millimetric extragalactic foreground
derived by Fixsen et~al. 1998  is given by
$I_{F98} \simeq 1.3 \times 10^{-5} [\nu/(c/0.01{\rm cm})]^{0.64} B_\nu(18.5{\rm K})$.
These results support a high redshift ($z \simeq 2.1-3.8$) active phase of 
star formation rate and dust reprocessing (Burigana et~al. 1997,
Sadat et~al. 2001). 
The levels of the sub-millimetric extragalactic foreground
theoretically predicted on the
basis of the number counts modelled by 
Franceschini et al. (1994) and revised by
Toffolatti et~al. (1998) or modelled by
Guiderdoni et~al. (1998) are in quite good agreement one each other
(within a factor $\sim 2$) and quite consistent
with the sub-millimetric extragalactic monopole as derived 
in the three above works, being respectively only just below or above it
(see also, e.g., De~Zotti et~al. 1999).
In addition, the subtraction of the isotropic residual in the FIRAS (LLSS) data 
as firstly modelled
by Fixsen et~al. 1996 in terms of a relatively steep 
spectral form, $I_{F96} \simeq G_0 (\nu/{\rm Hz})^\beta B_\nu(T_d)$,
being $G_0 \simeq 4.63 \times 10^{-29}$ (Fixsen, private communication),
$\beta =2$ the dust emissivity index and $B_\nu(T_d)$
the brightness of a blackbody at the dust temperature $T_d = 9$~K,
does not produce a significant lowering of 
the upper limits on spectral distortions, as discussed by 
Burigana et~al. 1997. In fact, they found that the allowed ranges for 
spectral distortion parameters do not change substantially by 
allowing for somewhat different astrophysical monopole shapes,
as in the case of a dust emissivity index varying in the range
$1.5 \lsim \beta \lsim 2.1$ consistent with 
the study of cosmic dust grains by Mennella et~al. 1998.

On the other hand, Battistelli et~al. (2000) recently reconsidered the absolute 
calibration of the FIRAS data on the basis of numerical simulations
of the external calibrator emissivity. 
As a variance with respect to the previous analysis by Fixsen et al. 1996,
they found an emissivity 
essentially constant within the 0.01\% at wavelengths 
$1~{\rm mm} \gsim \lambda \gsim 600 \mu{\rm m}$, where FIRAS 
sensitivity is particularly good, and decreasing 
with the wavelengths of up to about the 0.05\% in the range
$0.5~{\rm cm} \gsim \lambda \gsim 1~{\rm mm}$.
This translates into a significantly non flat shape of the 
monopole thermodynamic temperature, after 
the subtraction of the isotropic astrophysical foreground modelled
as discussed above. 
While a critical discussion of the FIRAS calibration from the experimental 
point of view is out of the scope of the present work,
we consider here the implications  
of this revised calibration.
Being the shape and the level of the sub-millimetric foreground quite 
well defined both from the observational and the theoretical point of view,
this revised FIRAS calibration,
significantly non flat at $\lambda \gsim 1$~mm,
should imply larger upper limits on CMB spectral distortion parameters, 
as suggested by Battistelli et~al. 2000,
(or even a possible detection of spectral distortions)  
or a presence of an astrophysical foreground possibly higher than that
derived from the extrapolation of the sub-millimetric 
foreground to wavelengths about or larger than 1~mm,
or, finally,  
a combination of these two effects.
We critically discuss these arguments in section~7.

\section{The data sets}\label{set_dati}

For the present study, we have extracted 
five different sets of measures. Four of them
have been chosen to take advantage from 
the very accurate informations from the FIRAS instrument aboard the COBE satellite. In 
particular, we considered that the statistical error associated to the 
measure at any channel of FIRAS is very small (0.02$\div$0.2~mK, 
Fixsen et al. 1996) and that the scale temperature at which the FIRAS 
data are set, have a systematic uncertainty of 2 mK at 95\% CL given by 
the calibration uncertainty (Mather et al. 1999). We 
analyze the impact of the calibration
uncertainty in the determination of the amount of the energy injected
in the cosmic background, when the FIRAS measures are considered together with
the data from ground and balloon experiments. Thus, we combine a set 
(see Table~1) 
of recent CMB spectrum data extracted from the complete database of 
CMB absolute temperature currently available at the 
different wavelengths (see, e.g., Salvaterra \& Burigana 2000) 
with the FIRAS data calibrated at the best-fit value as well as at the upper
and lower limit (at 95\% CL) of the temperature calibration.
Finally, we separately consider in the last set of data all the ground and 
balloon-borne measures reported in Table~1.
 
In the set of measures we do not include those from the COBRA experiment
nor those based on the analysis of the molecular lines, 
falling these experiments 
in the same frequency range of the much more accurate FIRAS measures.

\medskip

Summing up, we exploit five different data sets.

\medskip

Data $set \, 1)$: the FIRAS data alone (47 data points); 
the residuals reported in Table~4 of Fixsen et al. 1996
are added to a blackbody calibrated at the most recent value of 2.725~K 
(Mather et al. 1999)  and only the statistical errors channel by channel 
(see the uncertainties reported in Table~4 of Fixsen et al. 1996)
are taken into account. 
These data are completed by adding four points at 2.735 K (Mather et al. 1990)
in the range $1<\nu<2$ cm$^{-1}$ with a systematic error of 0.060 K, 
discarded in the following calibrations of the scale temperature.

\medskip

Data $set \, 2)$: the data of Table~1 of the last two decades
and the FIRAS data as above, i.e. calibrated at 2.725~K.

\medskip

Data $set \, 3)$: the data of Table~1 of the last two decades
and the FIRAS data as above but calibrated at the
lower calibration limit (2.723~K, 95\%~CL).

\medskip

Data $set \, 4)$: the data of Table~1 of the last two decades
and the FIRAS data as above but calibrated at the
upper calibration limit (2.727~K, 95\%~CL).

\medskip

Data $set \, 5)$: all the data of Table~1.

\medskip

%

In this way, we analyze in the three 
cases (2, 3, 4) the impact of the 
FIRAS data calibration in the determination 
of the amount of the energy possibly injected in the radiation field
without losing the important spectral shape
informations provided by the small statistical errors in the
channel by channel measures.
 
The comparison of the results obtained in the last case
with those obtained in the other cases allows 
to understand the different roles of ground
and balloon-borne measures and of the FIRAS data.
In some cases, we complete this analysis by considering 
other different combination of FIRAS data and long wavelength measures
(see sections~4.1.2 and 6).

\begin{table*}
\begin{center}
\begin{tabular}{lllll}
\hline
\hline
$\nu$ (GHz) & $\lambda$ (cm) & $T_{th}$ (K) & Error (K) & Reference\\
\hline
0.408 &  73.5 &  3.7 & 1.2 & Howell \& Shakeshaft 1967\\
0.610 &  49.1 & 3.7 & 1.2 & Howell \& Shakeshaft 1967\\
0.635 &  47.2 & 3.0 & 0.5 & Stankevich et al. 1970\\
1 & 30	&  2.5 & 0.3 & Pelyushenko \&  Stankevich 1969\\
1.42 & 21.2 & 3.2 & 1.0 & Penzias \& Wilson 1967\\
1.44 & 20.9 & 2.5 &  0.3 & Pelyushenko \& Stankevich 1969 \\
1.45 & 20.7 & 2.8 & 0.6 & Howell \& Shakeshaft 1966 \\
2 & 15 & 2.5 & 0.3 & Pelyushenko \& Stankevich 1969 \\
2.3 & 13.1 & 2.66 & 0.7 & Otoshi \& Stelzreid 1975 \\
4.08 & 7.35 & 3.5 & 1.0 & Penzias \& Wilson, 1965 \\
9.4 & 3.2 & 3.0 & 0.5 & Roll \& Wilkinson 1966 \\
9.4 &  3.2 & 2.69 & $+ 0.16/-0.21$ & Stokes et al. 1967 \\
19.0 & 1.58 & 2.78 & $+0.12/-0.17$ & Stokes et al. 1967 \\
20 & 1.5 & 2.0 & 0.4 & Welch et al 1967 \\
32.5 & 0.924 & 3.16 & 0.26 & Ewing et al. 1967 \\
35.0 & 0.856 & 2.56 & $+0.17/-0.22$ & Wilkinson 1967\\
37 & 0.82 & 2.9 & 0.7 & Puzanov et al. 1968 \\
83.8 & 0.358 & 2.4 & 0.7 & Kislyakov et al. 1971 \\
90 & 0.33 & 2.46 & $+0.40/-0.44$ & Boynton et al. 1968 \\
90 & 0.33 & 2.61 & 0.25 &  Millea et al. 1971\\
90 & 0.33 & 2.48 & 0.54 & Boynton \& Stokes 1974 \\
\hline
  0.6 & 50.0 & 3.0 &  1.2 &  Sironi et al. 1990\\
  0.82 &  36.6 & 2.7 &  1.6 &  Sironi et al. 1991\\
  1.28 & 23.3 & 3.45 & 0.78 & Raghunathan \& Subrahmanyan 2000\\ 
  1.4 &  21.3 &  2.11 &  0.38 &  Levin et al. 1988\\
  1.43 &   21 &  2.65 &  $+0.33/-0.30$ & Staggs et al. 1996a\\
  1.47 &   20.4 &  2.27 &  0.19 &  Bensadoun et al. 1993\\
  2 &   15 &  2.55 &  0.14 &  Bersanelli et al. 1994\\
  2.5 &   12 &  2.71 &  0.21 &  Sironi et al. 1991\\
  3.8 &    7.9 &  2.64 &  0.06 &  De Amici et al. 1991\\
  4.75 &    6.3 &  2.7 &  0.07 &  Mandolesi et al. 1986\\
  7.5 &    4.0 &  2.6 &  0.07 &  Kogut et al. 1990\\
  7.5 &   4.0 &  2.64 &  0.06 &  Levin et al. 1992\\
 10 &   3 &  2.62 &  0.058 &  Kogut et al. 1988\\
 10.7 &    2.8 &  2.730 &  0.014 &  Staggs et al. 1996b\\
 24.8 &    1.2 &  2.783 &  0.089 &  Johnson \& Wilkinson 1987\\
 33 &    0.909 &  2.81 &  0.12 &  De Amici et al. 1985\\
 90 &   0.33 & 2.60 &  0.09 &  Bersanelli et al. 1989\\
 90 &    0.33 &  2.712 &  0.020 &  Schuster 1993.\\
\hline
\end{tabular}
\end{center}
\caption{Measures of the absolute 
thermodynamic temperature of CMB spectrum considered in this work 
in addition to the COBE/FIRAS data. We separately report 
the data of the first two decades and of the last two decades 
since the CMB discovery.} 
\label{tab:rec}
\end{table*}


\section{Results}\label{sez:fit}

The fits are based on the $\chi^2$ analysis, 
carried out with a specific code based on the MINUIT package of the CERN library 
(http://cern.web.cern.ch/CERN/)
and on the 
set of subroutines and functions that implements the semi-analytical 
description of the CMB distorted spectra by Burigana et al. 1995. 
Our code allows to compare 
the CMB distorted spectrum models 
with the observational data without the necessity of interpolating 
frames of numerical solutions (as in Burigana et al. 1991b)
and to make the computation much more faster without 
any significant loss of accuracy, given the very good 
agreement between the semi-analytical expressions and the numerical solutions.

The CMB spectrum data are compared with the theoretical models by using the 
MINUT minimization routines 
(see the MINUITS/CERN documentation for further details).
The physical parameters that describe the distorted spectrum, can be 
set by choosing on which ones to have the fit.
A more detailed description of 
our code can be found in Burigana \& Salvaterra 2000.

We fit the different 
data sets with a  distorted spectrum 
for different values of the dimensionless
time parameter $y_h$ 
($y_h =$ 5, 4, 3, 2, 1, 0.5, 0.25, 0.1, 0.05, 0.025, 0.01 and $\ll1$)
in order to determine the value and the relative 
uncertainty of the present radiation temperature, $T_0$, 
and of the fractional energy, $\Delta \epsilon/\epsilon_i$.
[Of course, the redshift $z_h$ corresponding to a given value of $y_h$ 
decreases with the increase of baryon density (see eq.~2); 
for graphic purposes, we report in the
plots the exact values of $y_h$ and the power law
approximation $z_h(y_h)\simeq 4.94 \times 10^4 y_h^{0.477} \Ohat_b^{-0.473}$
(Burigana et al. 1991b) for the redshift.]

\subsection{Fits with a single energy injection}\label{un_heating}

In this section we present the results 
of the fits to the five data sets described in section~3 with
spectra distorted by a single energy injection. 

\bigskip

\subsubsection{Including FIRAS data}

The results obtained for the data $set \, 1)$ (FIRAS data only)
are shown in Fig.~1
for the full set of $y_h$ and 
the representative case $\Ohat_b=0.05$.
By comparing the results obtained for $\Ohat_b=0.01$ and 0.1
we find that the baryon density
does not influence the limits on $\Delta \epsilon/\epsilon_i$
derived from current data for cosmic epochs corresponding to 
the same dimensionless time $y_h$ of dissipation epoch, although 
the redshift corresponding to the same $y_h$ decreases with the baryon density.

The results are always compatible with null values
of the distortion parameters, the best fit values 
of $\Delta \epsilon/\epsilon_i$
being only just different from zero.
The $\chi^2$ best fit value does not change significantly with $y_h$ 
and $\Ohat_b$, being the $\chi^2$/DOF very close to unit; 
therefore this data do not show 
a favourite epoch for a possible (very small) energy injection nor
provide informations on the baryon density.
On the other hand, the limits on $\Delta \epsilon/\epsilon_i$ 
significantly depend on the epoch of the energy injection, being
about a factor two larger for early than for
late dissipation processes; this is clearly related to the 
range of frequencies observed by FIRAS.
 
In principle,  
the measures at centimetric and decimetric wavelengths 
could play a crucial role to investigate on the presence of early
distortions, due to the large decrease of the
CMB absolute temperature in the Rayleigh-Jeans region. 
The results found for the data $set \, 2)$ 
(recent data and FIRAS data calibrated at 2.725~K)
and $\Ohat_b=0.05$ are shown in 
Fig.~1; again, we find results essentially independent of the baryon density,
in spite the fact that the available ground
observations cover also the long wavelength spectral region where the amplitude of
possible early distortions strongly depend on $\Ohat_b$.
In general, as evident from the comparison with the results based on the 
data $set \, 1)$,
the ground and balloon data 
do not change significantly the constraints on energy dissipations 
with respect to the FIRAS measures alone, independently of the considered
energy injection epoch, because of their large error bars.

Even considering the full range of frequencies,
it is impossible to determine a favourite energy injection
epoch or a favourite baryon density value, as indicated by the $\chi^2$ values,
substantially constant with $y_h$ and $\Ohat_b$.
The $\chi^2$/DOF ($\sim 1.1$) is only just larger than that obtained
in the case of the data $set \, 1)$.
This weak increase of the $\chi^2$/DOF is due to the well known disagreement between 
the absolute temperature of the FIRAS data and the averaged temperature 
of the data at $\lambda \gsim 1$~cm.

We have analyzed also the impact of the 
calibration uncertainty (2~mK at 95\% CL, Mather et al. 1999) 
in the FIRAS data 
combined with the recent measures at $\lambda \gsim 1$~cm
(data $set \, 3)$ and data $set \, 4)$): 
we find a negligible impact on the fit results. 
By assuming the lowest FIRAS calibration 
only a small improvement (of $\sim 2\%$) in the $\chi^2$/DOF 
is found with respect to the case of the highest calibration, as expected
since the lower averaged temperature value at 
$\lambda \gsim 1$~cm than at $\lambda \lsim 1$~cm.
Thus, our analysis demonstrate that 
the current constraints on the energy possibly 
injected in the cosmic radiation field
are essentially set by the FIRAS measures alone
independently of the cosmic epoch.


%
\begin{figure*}
\epsfig{figure=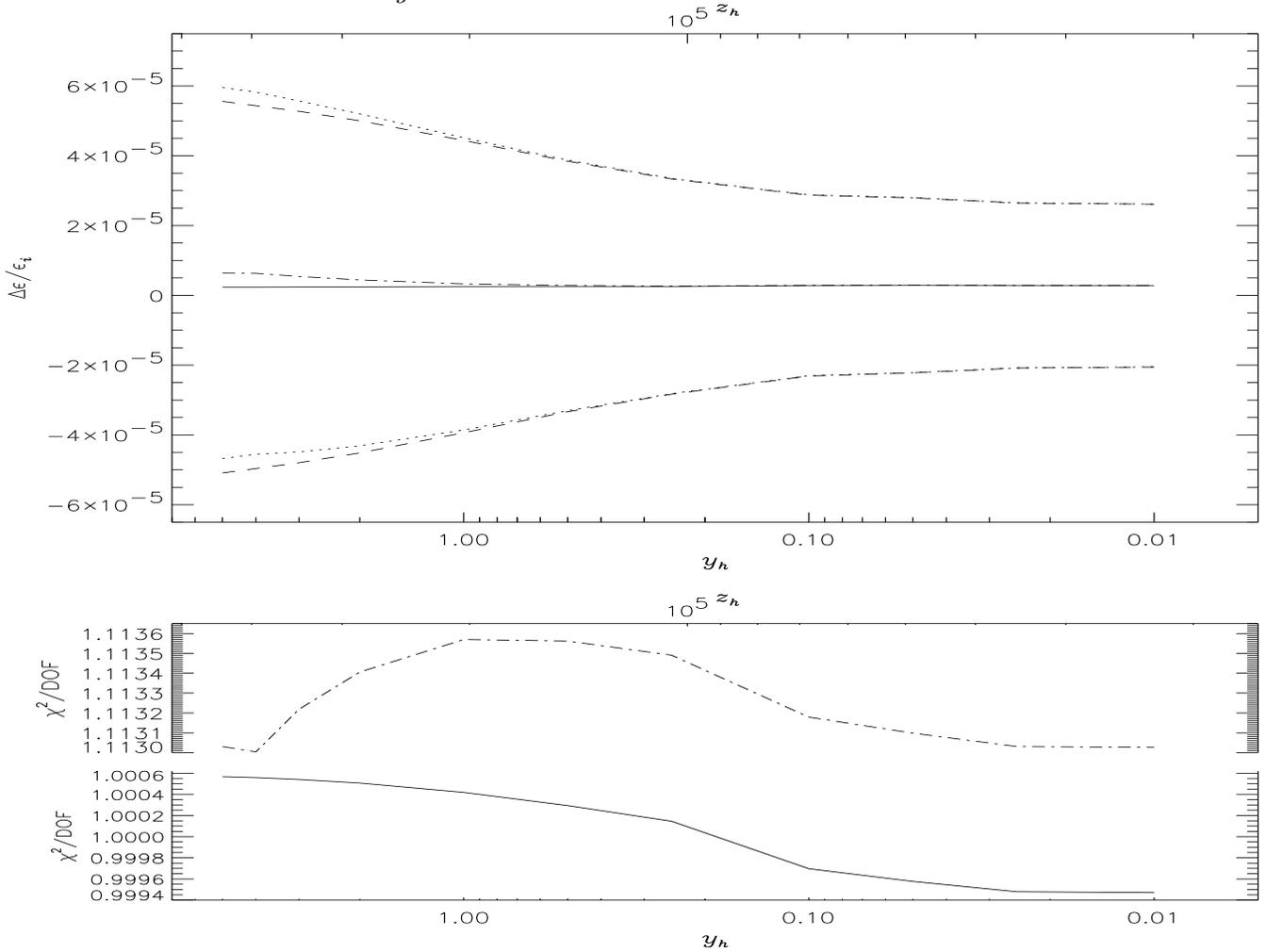,height=13cm,width=17cm}
\caption{Top panel: best fit and constraints at 95\% CL 
on $\Delta \epsilon/\epsilon_i$ 
as function of the energy dissipation epoch in the case 
on a single process occured in the thermal history of the universe.
Bottom panel: values of the $\chi^2$/DOF corresponding to the 
best fit curves.
We show here the results obtained from the 
data $set \, 1)$ (solid lines - best fit - and dashed lines - 
upper and lower limits)
and the data $set \, 2)$ (dot-dashed lines and dotted lines).
Note how current ground and balloon measures do not significantly modify 
the constraints on $\Delta \epsilon/\epsilon_i$ derived from 
the FIRAS data alone, while the $\chi^2$/DOF increases 
of about 0.11 [$\Ohat_b = 0.05$].
}
\end{figure*}

Finally, we report in Table~2 (rows 1--4) the values of
best-fit of the fractional injected energy,
$\Delta\epsilon/\epsilon_i$, in the case of a heating process 
at early ($y_h=5$, BE like spectrum) 
and late ($y_h\ll1$, Comptonizated spectrum) epochs,
with the corresponding uncertainties.



\begin{table*}
\begin{center}
\begin{tabular}{lccc}
\hline
\hline
  & & & \\
\multicolumn{1}{l}{Data set} & \multicolumn{2}{c}{$(\Delta\epsilon/\epsilon_i)/10^{-5}$} & $y_B/10^{-5}$ \\
\cline{2-3} \\
          & heating at & heating at & free-free dist. at \\
            & $y_h=5$    &  $y_h\ll1$    &  $y_h\ll1$    \\
\hline
 & & & \\ 
$set \, 1)$: FIRAS & $0.23\pm5.33$ & $0.28\pm2.33$ & $2.72\pm9.20$ \\
 & & & \\ 
$set \, 2)$: Table~1 (1985-2000) $+$ FIRAS (2.725~K) & $0.64\pm5.32$ & $0.28\pm2.33$ & $-4.55\pm4.24$ \\
 & & & \\ 
$set \, 3)$: Table~1 (1985-2000) $+$ FIRAS (2.723~K) & $0.63\pm5.33$ & $0.29\pm2.34$ & $-4.46\pm4.25$ \\
 & & & \\ 
$set \, 4)$: Table~1 (1985-2000) $+$ FIRAS (2.727~K) & $0.66\pm5.31$ & $0.29\pm2.33$ & $-4.64\pm4.23$ \\
 & & & \\ 
$set \, 5)$: Table~1 & $(1.36\pm1.32)\times 10^2 $ 
& $[-0.14 (+1.90/-1.47)] \times 10^4 $ & $-0.63\pm2.44$ \\
 & & & \\ 
\hline
\end{tabular}
\end{center}
\caption{Results on the energy injected 
at $y_h=5$ and at $y_h\ll1$ and on the free-free distortion parameter $y_B$.
Fits to the different data sets, errors at 95\% CL.
We jointly fit two parameters, $T_0$ and the $(\Delta\epsilon/\epsilon_i)$ referring to 
the early or late process or $y_B$, by assuming null values for the other
distortion parameters [$\Ohat_b=0.05$].}
\label{tab:fit:0.05}
\end{table*}


\subsubsection{Neglecting FIRAS data}

The results obtained for the data $set \, 5)$ (all the data of Table~1 without
including FIRAS data)
are shown in Fig.~2 for the full set of $y_h$ and 
the representative case $\Ohat_b=0.05$.

In this case the best fit values of $\Delta \epsilon/\epsilon_i$
for early processes ($y_h \gsim 2$),
although compatible with a BB spectrum at about $2\sigma$ level,
are significantly (at about $1\sigma$ level) positive, ranging from 
$\simeq 1.4 \times 10^{-3}$ to  $\simeq 2 \times 10^{-3}$ 
for $5 \gsim y_h \gsim 2$.
On the contrary, 
from this long wavelength set of data it is no longer possible 
to tightly constrain dissipation processes at $y_h \lsim 1$.

We find a $\chi^2$ best fit value slightly decreasing with $y_h$.
We find also a weak, but not completely negligible, dependence 
of $\Delta\epsilon/\epsilon_i$ on $\Ohat_b$:
for example, for $\Ohat_b=0.01$ ($\Ohat_b=0.1$)
the (95\% CL) upper limit on 
$\Delta\epsilon/\epsilon_i$ decreases (increases) of about 1.5\%
(14\%) with respect to upper limit obtained for $\Ohat_b=0.05$.

We report in Table~2 (row 5) the values of
best-fit of the fractional injected energy,
$\Delta\epsilon/\epsilon_i$, in the case of a heating process 
at early ($y_h=5$, BE like spectrum) 
and late ($y_h\ll1$, Comptonizated spectrum) epochs,
with the corresponding uncertainties at 95\% CL. 

Finally, we note that these results are mainly driven by the 
observations of the last two  
decades.
In fact, by considering only these measures 
we obtain
$(\Delta\epsilon/\epsilon_i)/10^{-5} = (1.57\pm1.45)\times 10^2$ and
$(\Delta\epsilon/\epsilon_i)/10^{-5}  = [-0.10 (+1.93/-1.49)] \times 10^4$ respectively
in the case of an early ($y_h=5$, BE like spectrum) 
and a late ($y_h\ll1$, Comptonizated spectrum) process
(errors at 95\% CL).

\begin{figure*}
\epsfig{figure=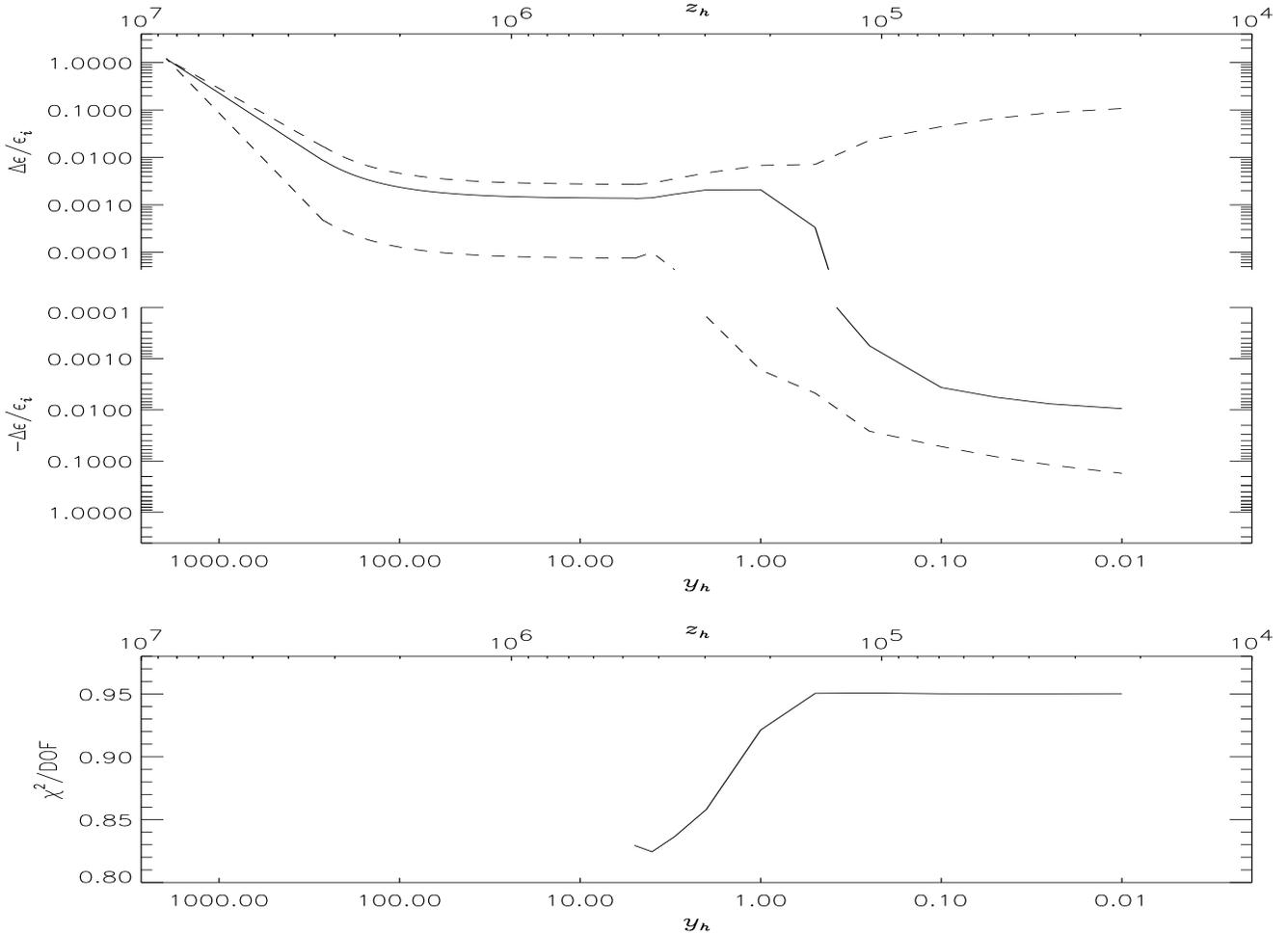,height=13cm,width=17cm}
\caption{The same as in Fig.~1, but by exploiting 
all the data of Table~1 and neglecting the FIRAS data.
We report here also the constraints on the energy 
exchange at very high redshifts, as described in  
section~5, for comparison with the results shown in Fig.~4.
Of course, 
the constraints on $\Delta \epsilon/\epsilon_i$ obtained 
by neglecting FIRAS data are significantly
relaxed at each cosmic epoch [$\Ohat_b = 0.05$].
}
\end{figure*}

\subsection{Fits with two energy injections}\label{2heating}

We analyse here the constraints set by the available measures
when we take into account the possibility that two heating processes 
could have distorted the CMB spectrum at different epochs, early or intermediate
for the former and late for the latter. 
More explicitly, we obtain the limits on the amount of the first 
energy injection for each value of $y_h$ (in the range $5\geq y_h \geq 0.01$)
under the hypothesis of a possible existence of a second late heating 
(at $y_h\ll 1$) and 
on the amount of the second late energy injection (at $y_h\ll 1$) 
under the hypothesis of a possible existence 
of an earlier energy dissipation occurring at different values of $y_h$
(in the range $5\geq y_h \geq 0.01$).
So far, we extend the analysis of Nordberg \& Smoot 1998 
which considered only the case of an early dissipation at $y_h =5$
combined with a late one at $y_h \ll 1$.
Our results are then clearly comparable 
to those obtained in the section~4.1:
we expect that, 
by including 
the possibility of two heating processes at different cosmic epochs,
the constraints on $\Delta \epsilon/\epsilon_i$,
both for early and late processes, 
are relaxed 
with respect to the case in which a single heating 
in the thermal history of the universe is considered.

Of course, it is particularly interesting in this context to exploit 
the whole frequency range 
of the recent CMB spectrum measures [i.e. the data 
$set \, 2)$, $3)$, and $4)$].
For comparison, in the exploitation of the data $set \, 1)$ 
we circumscribe our joint analysis just to the simple circumstance of 
two processes at $y_h=5$ and $y_h\ll1$ (see Table~3),
while neglecting FIRAS data 
makes this analysis meaningless, as evident from 
the results of section~4.1.2 in the case of late distortions.

We report here in detail the results obtained 
by exploiting the data $set \, 2)$ for $\Ohat_b=0.05$ (see Fig.~3); 
again, the constraints
on the energy dissipated at different epochs are 
essentially independent of the baryon density when expressed
in terms of the dimensionless time $y_h$.
The top panel of Fig.~3 
shows the best fit value of the fractional energy exchanged 
in the plasma as a function of the cosmic epoch 
(for $5 \gsim y_h \gsim 0.01$)
and its upper and lower limits at 95\% CL when 
we allow for a possible later dissipation process
($y_h \ll 1$). 
The middle panel 
reports the best fit value of the fractional energy exchanged 
in the plasma at $y_h \ll 1$ 
and its upper and lower limits 
when we allow for 
a possible earlier dissipation process
(occurring at $5 \gsim y_h \gsim 0.01$).
The bottom panel 
gives the $\chi^2$/DOF corresponding to the best fit model
obtained by allowing for a late dissipation and an earlier 
dissipation at a certain epoch $y_h$ ($5 \gsim y_h \gsim 0.01$).

\begin{figure*}
\epsfig{figure=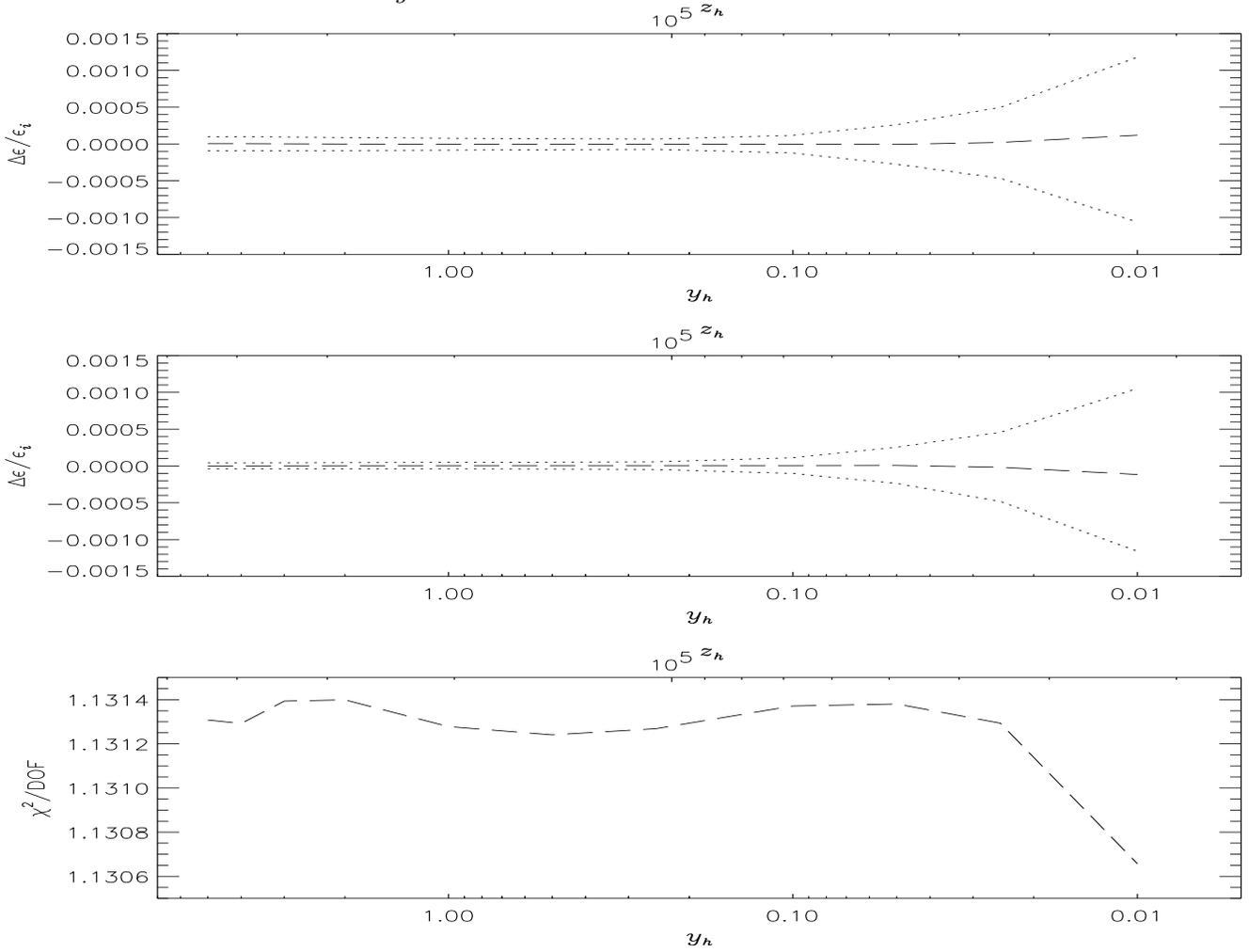,height=13cm,width=17cm}
\caption{Top panel: best fit (long dashes)
and constraints (dotted lines) at 95\% CL 
on the earlier ($y_h \gsim 0.01$) energy exchange, $\Delta \epsilon/\epsilon_i$, 
as function of the dissipation epoch when we allow also for 
a late ($y_h \ll 1$) energy exchange 
in the thermal history of the universe.
Middle panel: best fit and constraints at 95\% CL 
on the late ($y_h \ll 1$) energy exchange, $\Delta \epsilon/\epsilon_i$, 
as function of the dissipation epoch 
of an earlier ($y_h \gsim 0.01$) energy exchange occurring at a given epoch 
in the thermal history of the universe.
Bottom panel: values of the $\chi^2$/DOF corresponding to the 
best fit curves.
We show here the results obtained by exploiting the 
data $set \, 2)$ [$\Ohat_b = 0.05$].
}
\end{figure*}

We note here that, in general, the joint analysis of two 
dissipation processes results to be meaningless  
for earlier processes occurring at $y_h<0.1$ (when  
the limits on the amount of injected energy became 
very relaxed), because the imprints 
produced by a positive (negative) 
earlier distortion at any $y_h<0.1$ can be partially compensated by those
produced by a later negative (positive) distortion at $y_h\ll1$ (and this 
cancellation effect increases for an earlier distortion 
occurring at smaller and smaller $y_h$), owing to the similarity 
of the distorted spectral shapes at small $y_h$.

As shown by the values reported in 
Table~3 for the case of a joint analysis of a heating process at $y_h=5$ and 
one at $y_h\ll1$, the limits
on $\Delta \epsilon/\epsilon_i$ 
are relaxed by a factor $\sim 2$ 
with respect to the case of in which a single energy injection 
in the thermal history is considered,
both for early and late dissipation processes 
(see Table~2 for comparison).

Finally, the exact FIRAS calibration returns to be not crucial 
in the exploitation of current data and 
the $\chi^2$/DOF value is substantially constant,
and close to unit, for the 
different values of $y_h$ and $\Ohat_b$.


\begin{table*}
\begin{center}
\begin{tabular}{lcc}
\hline
\hline
  & & \\
\multicolumn{1}{l}{Data set} & \multicolumn{2}{c}{$(\Delta\epsilon/\epsilon_i)/10^{-5}$} \\
\cline{2-3} \\
          & heating at & heating at \\
            & $y_h=5$    & $y_h\ll1$ \\
\hline
 & & \\
$set \, 1)$: FIRAS & $0.95\pm9.67$ & $0.61\pm4.18$ \\
 & & \\
$set \, 2)$: Table~1 (1985-2000) $+$ FIRAS (2.725~K) & $0.37\pm9.62$ & $0.14\pm4.17$ \\
 & & \\
$set \, 3)$: Table~1 (1985-2000) $+$ FIRAS (2.723~K) & $0.27\pm9.65$ & $0.19\pm4.19$ \\
 & & \\
$set \, 4)$: Table~1 (1985-2000) $+$ FIRAS (2.727~K) & $0.42\pm9.61$ & $0.12\pm4.17$ \\
 & & \\
\hline
\end{tabular}
\end{center}
\caption{Results on the energy injected at $y_h=5$ and $y_h\ll1$ 
when these two dissipation processes are jointly considered.
Fits to the different data sets, errors at 95\% CL.
We jointly fit three parameters: $T_0$ and the two values of 
$(\Delta\epsilon/\epsilon_i)$ referring to 
the early and late process [$\Ohat_b=0.05$].}
\label{tab:fit:2h_0.05}
\end{table*}


\section{Constraints on energy injections at very high redshifts}
\label{evo_attuale}

For $z>z_1$ (i.e. $y_h > 5$) the Compton
scattering is able, after an energy injection, to restore the kinetic 
equilibrium between matter and radiation yielding a BE spectrum
on timescales smaller than the expansion time, while
radiative Compton and Bremsstrahlung work to restore the thermodynamic
equilibrium yielding a BB spectrum.
Thus, a larger amount of energy would have been needed to yield the same 
observational effect produced by a dissipation processes at $\sim z_1$. 
The analytic approximations of Burigana et al. 1991b 
of the numerical computations carried out by Burigana et al. 1991a,b 
permits us to extend the limits on $\Delta\epsilon/\epsilon_i$ at 
$z_h>z_1$ with good accuracy without the necessity of numerical
integrations of the chemical potential and electron temperature
evolution equations. 
We have reported here, for simplicity, 
only the results for $\Delta \epsilon /\epsilon_i (y_h)$
in the case $\Ohat_b=0.05$, 
but analogous constraints can be easily derived for any 
value of $\Ohat_b$.
 
The limits on $\Delta\epsilon/\epsilon_i$ 
at 95\% CL obtained from the 
accurate measures of FIRAS data alone [data $set \, 1)$] and 
from the recent measures from ground 
and balloon combined with the FIRAS data calibrated at 2.725~K 
[data $set \, 2)$]
are very close one each other (see Fig.~4), in strict analogy
with the corresponding limits on the amount of energy injected at $z=z_1$.
The FIRAS data significantly constrain
the value of $\Delta\epsilon/\epsilon_i$ also at $z>z_1$;
of course, a very large energy dissipation
at $z \approx z_{therm}$ can not be excluded from CMB spectrum observations.
The limits on the energy possibly 
injected in the radiation field 
at $z \gsim z_{therm}$ can be set by primordial nucleosynthesis
analyses.

We obtain also the limits on  $\Delta\epsilon/\epsilon_i$ at high $z$ 
by allowing for a  second heating possibly occurred at low  $z$
(see again Fig.~4).
In this case, the limits
at 95\% CL on the amount of the energy injected at $z=z_1$, 
relaxed compared to the case of a single heating in the thermal history 
of the universe,
allow for larger energy dissipations (by a factor up to $\sim 2$), 
particularly at $z_1 \lsim z \lsim z_{therm}/2$.

Finally, by neglecting FIRAS data and keeping all the data of Table~1 
[data $set \, 5)$],
the constraints on $\Delta \epsilon/\epsilon_i$ 
result to be much less stringent at each cosmic epoch, as shown
in Fig.~2 for the case of a single energy exchange 
in the thermal history of the universe.

\begin{figure*}
\epsfig{figure=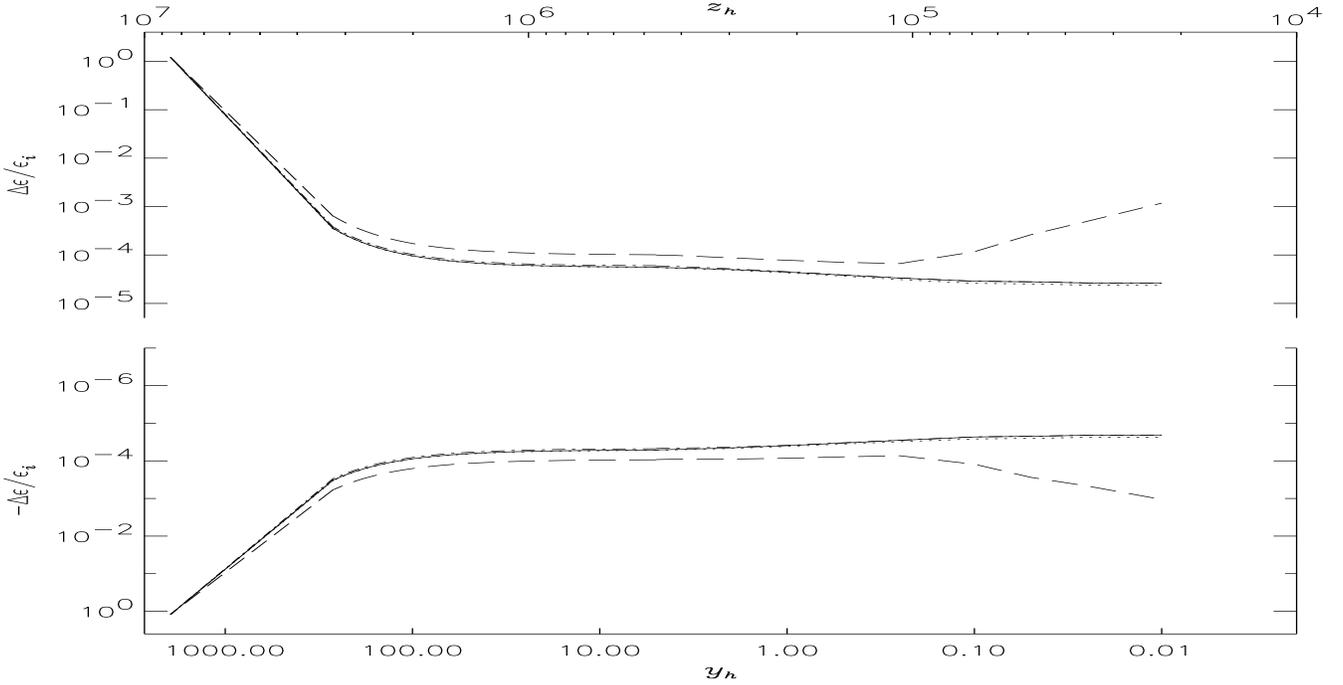,height=10cm,width=17cm}
\caption{Extension to very high redshifts 
of the constraints at 95\% CL 
on the energy exchange, $\Delta \epsilon/\epsilon_i$, 
as function of the dissipation epoch.
We consider here several cases:
the constraints on a single energy injection 
in the thermal history of the universe by exploiting 
the data $set \, 1)$ (solid lines) and the data 
$set \, 2)$ (dot-dashed lines);
the constraints on the earlier ($y_h \gsim 0.01$) energy exchange, 
$\Delta \epsilon/\epsilon_i$, 
when we allow also for a late ($y_h \ll 1$) energy exchange 
in the thermal history of the universe and exploit 
the data $set \, 2)$ (dashed lines).
We report also the constraints in the case 
of a single energy injection in the thermal history of the universe
when we exploit the FIRAS data alone but calibrated according to 
Battistelli et al.~(2000) by adding the ``experimental'' astrophysical 
monopole, $I_{F96}$, 
derived by Fixsen et al. (1996) and subtracting the ``theoretical'' 
astrophysical monopole of the best fit obtained assuming a power law plus a 
dust emission law (dotted lines) as in Table~14.
Exploiting different data sets, including FIRAS data, does not change 
significantly the result, as evident from the superposition
of the curves referring to the different cases;
this is true also in the case 
of the calibration by Battistelli et al.~(2000) provided that
the astrophysical component is properly modelled (see section~7).
The only relevant difference appears when we consider
a single or a double energy exchange in the thermal history of the universe
[$\Ohat_b = 0.05$].} 
\end{figure*}

\section{Free-free distortions}

As discussed in section~3.2, a proper joint analysis of Comptonization
and free-free distortions requires the specification of the 
kind of considered thermal history at late epochs.

On the other hand, it is also interesting to evaluate the constraints 
set by current observations on free-free distortions and the 
possible impact of 
free-free distortions on the estimation  of the other distortion 
parameters.

In the last column of Table~2 we report the best fit results and the 
limits on $y_B$ derived by performing the fit to  the different 
data sets for $T_0$ and $y_B$ under the assumption of negligible
energy exchanges ($\Delta \epsilon /\epsilon_i = 0$ at any $y_h$).
By varying the data from the $ set\, 1)$
to the $set \, 4)$, the $\chi^2$/DOF assumes the values $\simeq 0.993$, 1.040,
1.045 and 1.035. It is also evident from the table the 
inconsistency at $\simeq 1\sigma$ level 
between the results based on FIRAS data alone and on the data set 
including also the recent measures at longer wavelengths.
We note also that, on the contrary, the limits on $y_B$ based
on all the data of Table~1 alone are consistent with those based
on FIRAS data alone. The limits on free-free distortions based on
all the data of Table~1 are in fact different from those derived
considering only the data of the two last decades
($y_B/10^{-5} = -5.49\pm4.92$). 
This is mainly due to the 
measures by Levin et~al. 1988 and Bensadoun et~al. 1993; neglecting them,
from the data of the two last decades we find $y_B/10^{-5} = -2.23\pm6.10$.
[Errors at 95\% CL].


The above inconsistency is again present by allowing also for energy exchanges
both at high ($y_h = 5$) and low ($y_h \ll 1$) redshifts (see the last column
of Table~4). In Table~4 note also that by 
exploiting the full frequency range of the recent measures
to simultaneously fit $T_0$, $\Delta \epsilon /\epsilon_i (y_h=5)$,
$\Delta \epsilon /\epsilon_i (y_h \ll1)$ and $y_B$ 
the signs of $\Delta \epsilon /\epsilon_i (y_h \ll1)$ and $y_B$ 
are close to be inconsistent at $\simeq 1\sigma$ level. 
Although the value of the $\chi^2$/DOF ($\simeq 1.062$)
represents an improvement with respect to 
the fit reported in the second row of Table~3
($\chi^2$/DOF~$\simeq 1.131$),
it is difficult from a physical point of view
to explain this inconsistency
(a precise fine tuning in the late thermal history
is required to produce $\Delta \epsilon /\epsilon_i (y_h \ll1)$ and $y_B$ 
with different signs).
Again, we note that by using FIRAS data together
with the recent data of Table~1 without 
the measures by Levin et~al. 1988 and Bensadoun et~al. 1993
or together with all the data of Table~1 
to jointly fit $T_0$, $\Delta\epsilon/\epsilon_i (y_h =5)$,
$\Delta\epsilon/\epsilon_i (y_h \ll1)$ and $y_B$
this inconsistency 
disappears (we find, respectively,
$(\Delta\epsilon/\epsilon_i (y_h =5))/10^{-5} = -1.75\pm10.83$, 
$(\Delta\epsilon/\epsilon_i (y_h \ll1))/10^{-5}  = 0.79\pm4.40$, 
$y_B/10^{-5} = -1.84\pm5.55$
and
$(\Delta\epsilon/\epsilon_i (y_h =5))/10^{-5} = -0.29\pm9.92$, 
$(\Delta\epsilon/\epsilon_i (y_h \ll1))/10^{-5}  = 0.34\pm4.23$, 
$y_B/10^{-5} = -0.71\pm2.37$, errors at 95\% CL).

Clearly, more precise observations at long wavelengths are required
for a proper evaluation of free-free distortions.

Finally, we note that, in spite of the above inconsistencies,
the constraints on $\Delta \epsilon /\epsilon_i (y_h=5)$ and on
$\Delta \epsilon /\epsilon_i (y_h \ll1)$ are not particularly modified
with respect to the results shown in Table~3, where possible free-free 
distortions are 
neglected~\footnote{
For the data $set \, 1)$, which not presents the above inconsistency, 
the upper limits on $\Delta \epsilon /\epsilon_i (y_h \ll1)$ are 
only just modified, whereas a positive free-free distortion 
can be clearly compensated in part  by 
a somewhat larger energy injection at high redshift
(compare the first row of Table~3 -- $\chi^2$/DOF~$\simeq 1.021$ --
and of Table~4 -- $\chi^2$/DOF~$\simeq 1.029$).}.
This is particularly important, because the constraints
on many classes of astrophysical processes are typically based on energetic
arguments (see, e.g., Platania et~al. 2002 for a recent
application).

\begin{table*}
\begin{center}
\begin{tabular}{lccc}
\hline
\hline
  & & & \\
\multicolumn{1}{l}{Data set} & \multicolumn{2}{c}{$(\Delta\epsilon/\epsilon_i)/10^{-5}$} & $y_B/10^{-5}$  \\
\cline{2-3} \\
          & heating at & heating at & free-free dist. at \\
            & $y_h=5$    & $y_h\ll1$ & $y_h\ll1$ \\
\hline
 & & & \\
$set \, 1)$: FIRAS & $4.14\pm15.61$ & $-0.85\pm5.46$ & $6.35\pm15.31$ \\
 & & & \\
$set \, 2)$: Table~1 (1985-2000) $+$ FIRAS (2.725~K) & $4.65\pm10.56$ & $1.63\pm4.36$ & $-5.40\pm4.69$ \\
 & & & \\
\hline
\end{tabular}
\end{center}
\caption{Results on the energy injected at $y_h=5$ and $y_h\ll1$ 
and on the free-free distortion parameter when these three 
types of distortions are jointly considered.
Fits to the different data sets, errors at 95\% CL.
We jointly fit four parameters: $T_0$, the two values of 
$(\Delta\epsilon/\epsilon_i)$ referring to 
the early and late process and $y_B$ [$\Ohat_b=0.05$].}
\label{tab:fit:2h_0.05}
\end{table*}

\section{Implications of a revised calibration of FIRAS data}

The calibration of the FIRAS data as function of the frequency
(i.e. its spectral shape), particularly in the Low (LLSS) frequency bands,
is crucial to probe CMB spectral distortions because of the extremely 
good sensitivity of FIRAS ($\sim 0.1$~mK, see Fig.~5) at each frequency channel.
In the previous sections we have assumed the FIRAS calibration 
derived by the COBE/FIRAS team as described by Fixsen et~al. 1994, 
Fixsen et~al. 1996 and revised by Mather et~al. 1999. 
According to the authors,
the external calibrator shows a very low reflectance, $\rho$, 
less than $3\times 10^{-5}$ at $\lambda = 1$~cm and decreasing with the frequency, 
and then an emissivity function, $e = 1-\rho$, very close to unit 
at each wavelength. 
Battistelli et~al. (2000) have recently reanalysed the emissivity 
of the calibrator by using numerical simulations based on 
Ray-Tracing techniques. They find that the relatively small
dimension of the inner part of the V-groove could distort the emission 
spectrum when it becomes comparable with the wavelength.
In this condition the diffraction may be relevant 
in the final reflection. The authors studied this problem 
by using a waveguide model (``wave radial waveguide'') 
to compute the emissivity function of the external calibrator
under various diffraction configurations.

\subsection{The recalibrated data set}

According to the results reported in Fig.~12 of Battistelli et~al. (2000),
the spectrum of the FIRAS calibrator
is given by a blackbody (assumed here at a temperature of 2.725~K,
according to Mather et~al. 1999) multiplied by a frequency dependent
emissivity function, $e$, very close to unit at the highest frequencies but
significantly decreasing with the wavelength for $\lambda \gsim 0.1$~cm.
As stressed by the authors, a completely exhaustive analysis of 
FIRAS calibration is a very difficult task. The calibration spectral shape
may be conservatively considered as varying from the case of constant
emissivity (i.e., a pure blackbody shape) to a case of 
maximum deviation from a blackbody shape 
which shows an emissivity decrease of about 
0.07\% at $\lambda \simeq 0.5$~cm. In the former case 
the measures are substantially equivalent to the data $set \, 1)$ described
in section~3. In the latter case, the revised calibration implies 
a maximum deviation from the Planckian spectrum. 
We consider here the cosmological and astrophysical implications 
of the calibrator emissivity law 
given by Table~1 of Battistelli et~al. 2000 
which ``represents the conservation of the 
transmitted flux inside the radiator''.
We construct then a new set of CMB absolute temperature 
(referred as R-FIRAS in the following tables and figure captions)
by adding this revised calibration spectrum 
to the residuals of Table~4 of Fixsen et~al. 1996
by assuming the same statistical uncertainties.
As evident from Fig.~5, this revised calibration 
is substantially equivalent to that of Mather et al. 1999 
for $\lambda \lsim 0.1$~cm, but 
implies a decrement of the CMB absolute temperature
of about 1~mK from $\simeq 0.1$~cm to $\simeq 0.5$~cm.
This calls for an accuracy of future CMB spectrum measurements 
about $\sim 1$~cm better than $\sim 1$~mK 
(both in sensitivity and calibration) to clearly distinguish 
between the two FIRAS data calibration shapes.

In sections~$7.2 \div 7.5$  we will analyse
the FIRAS data so revised without 
resorting to particular cosmological or astrophysical considerations.
We will include them in the discussion of section~7.6. 

\begin{figure*}
\epsfig{figure=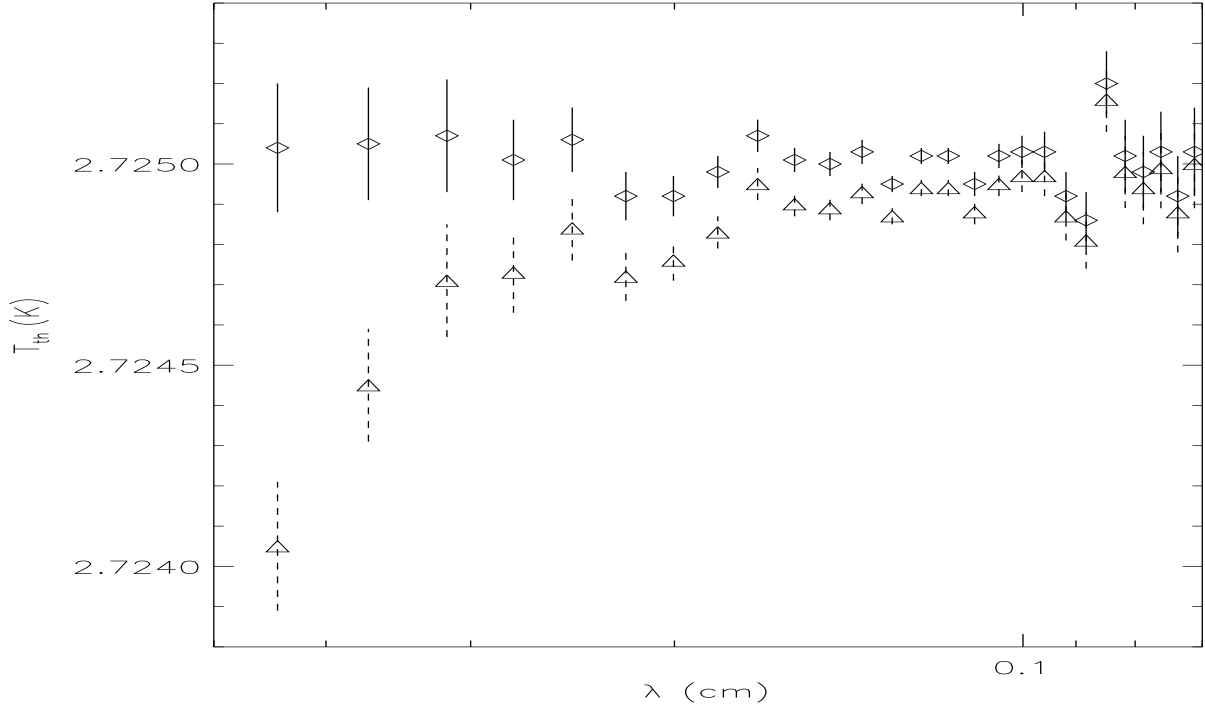,height=10cm,width=17cm}
\caption{Comparison between the CMB absolute 
thermodynamic temperatures derived from FIRAS data calibrated 
according to Mather et~al. 1999 (diamonds) and to Battistelli 
et~al. (2000) (triangles) [see also the text].}
\end{figure*}

\subsection{Interpretation in terms of pure CMB spectral distortions}

We tried to explain this revised FIRAS data set in terms of pure CMB
spectral distortions. Our results are summarized in Tables~5, 6 and 7, where
we consider separately the case of an early heating process, a late heating process
and a free-free distortion with negligible energy exchange, the 
joint effect of an early and a late dissipation process, and, finally, 
the joint effect of an early and a late dissipation process associated 
to a non negligible free-free distortion, respectively.
As evident from the fit results, these revised FIRAS data support 
the existence an early dissipation process 
with $\Delta\epsilon/\epsilon_i \sim (0.5 \div 4) \times 10^{-4}$.
Comptonization distortions only marginally improve the fit 
(by reducing of only about the 5\% the $\chi^2$/DOF for a late cooling process,
compare Tables~6 and 7 with the second column of Table~5).
Adding free-free distortions does not improve the fit 
(compare Table~7 with Table~6). As discussed in section~6, precise 
long wavelength observations are required to evaluate free-free
distortions; we neglect them in the following considerations. 

We observe that, in any case, by considering only CMB spectral distortions
the value of the $\chi^2$/DOF is always larger than $\simeq 1.2$, even 
in this favourite case in which a single set of measurements in considered.

\begin{table*}
\begin{center}
\begin{tabular}{lccc}
\hline
\hline
  & & & \\
\multicolumn{1}{l}{Data set} & \multicolumn{2}{c}{$(\Delta\epsilon/\epsilon_i)/10^{-5}$} & $y_B/10^{-5}$ \\
\cline{2-3} \\
          & heating at & heating at & free-free dist. at \\
            & $y_h=5$    &  $y_h\ll1$    &  $y_h\ll1$    \\
\hline
 & & & \\ 
R-FIRAS & $20.15\pm5.32$ & $6.07\pm2.30$ & $-28.71\pm9.19$ \\
 & & & \\ 
$\Delta T_0({\rm K})/10^{-4}$ & $-1.16\pm0.16$ & $-0.71\pm0.17$ & $-0.89\pm0.15$ \\ 
 & & & \\
$\chi^2$/DOF &  1.248     &   1.906    &   1.657   \\
 & & & \\ 
\hline
\end{tabular}
\end{center}
\caption{Results on the energy injected 
at $y_h=5$ and at $y_h\ll1$ and on the free-free distortion parameter $y_B$.
Fits to the FIRAS data with the revised calibration, errors at 95\% CL.
We jointly fit two parameters, $T_0$ and the $(\Delta\epsilon/\epsilon_i)$ referring to 
the early or the late process or the parameter $y_B$, by assuming null values for the other
distortion parameters [$\Ohat_b=0.05$].}
\label{tab:fit:0.05}
\end{table*}
\begin{table*}
\begin{center}
\begin{tabular}{lcccc}
\hline
\hline
  & & & & \\
\multicolumn{1}{l}{Data set} & \multicolumn{2}{c}{$(\Delta\epsilon/\epsilon_i)/10^{-5}$} & $\Delta T_0({\rm K})/10^{-4}$ & $\chi^2$/DOF \\
\cline{2-3} \\
          & heating at & heating at & &\\
            & $y_h=5$    &  $y_h\ll1$ & &\\
\hline
 & & & & \\ 
R-FIRAS & $27.83\pm9.66$ & $-3.99\pm4.18$ & $-1.40\pm0.29$ & 1.194\\
 & & & & \\ 
\hline
\end{tabular}
\end{center}
\caption{Results on the energy injected 
at $y_h=5$ and at $y_h\ll1$ by assuming a null free-free distortion parameter $y_B$.
Fits to the FIRAS data with the revised calibration, errors at 95\% CL.
We jointly fit three parameters: $T_0$ and the two $(\Delta\epsilon/\epsilon_i)$ values referring to 
the early and the late process [$\Ohat_b=0.05$].}
\label{tab:fit:0.05}
\end{table*}
\begin{table*}
\begin{center}
\begin{tabular}{lccccc}
\hline
\hline
  & & & & & \\
\multicolumn{1}{l}{Data set} & \multicolumn{2}{c}{$(\Delta\epsilon/\epsilon_i)/10^{-5}$} & 
$y_B/10^{-5}$ & $\Delta T_0({\rm K})/10^{-4}$ & $\chi^2$/DOF \\
\cline{2-3} \\
          & heating at & heating at & free-free dist. at & & \\
            & $y_h=5$    &  $y_h\ll1$    &  $y_h\ll1$   &  & \\
\hline
 & & & & & \\ 
R-FIRAS & $21.86\pm15.62$ & $-2.27\pm5.46$ & $-7.45\pm15.30$ & $-1.25\pm0.42$ & 1.200\\
 & & & & & \\ 
\hline
\end{tabular}
\end{center}
\caption{Results on the energy injected 
at $y_h=5$ and at $y_h\ll1$ and on the free-free distortion parameter $y_B$.
Fits to the FIRAS data with the revised calibration, errors at 95\% CL.
We jointly fit four parameters: $T_0$, the two $(\Delta\epsilon/\epsilon_i)$ values referring to 
the early and the late process and $y_B$ [$\Ohat_b=0.05$].}
\label{tab:fit:0.05}
\end{table*}

\subsection{Interpretation in terms of millimetric astrophysical foreground}

Given the relatively high $\chi^2$/DOF obtained in the previous section
when the revised FIRAS data are interpreted in terms of pure CMB distortions,
we investigate here if it is possible to better explain them 
by supposing the existence of a monopole of astrophysical nature not
subtracted in the data reduction with a significant contribution at
millimetric wavelengths.
We try a fit in terms of a power law,
$I_{PL} = {\rm k} (\lambda / {\rm cm})^{-\alpha} \,$,
and in terms of a dust emission law
approximated by a modified blackbody,
$I_{D} = {\rm k}_d (\lambda / {\rm cm})^{-\beta -3} / [{\rm exp}(h\nu/kT_d)-1] \,$;
we report the following fit results in terms 
of ${\rm log}\, {\rm k}$ and ${\rm log} \, {\rm k}_d$ 
with ${\rm k}$ and ${\rm k_d}$ expressed in units of  
${\rm erg} \, {\rm cm}^{-2} {\rm sec}^{-1} {\rm sr}^{-1} {\rm Hz}^{-1}$.

The fit results are reported in Tables~8 and 9. 
As evident, a single modified blackbody component provides a somewhat 
better fit than a combination of two or three CMB spectral distortions.
Although described by three parameters, as in the case of Table~7,
the interpretation of these data in terms of a single dust emission law
at a temperature only just above that of the CMB 
seems to be easier than a proper combination of two 
dissipation processes at different cosmic epochs, 
but in any case the $\chi^2$/DOF, of about 1.2, 
still remains significantly higher than unit.

\begin{table*}
\begin{center}
\begin{tabular}{lcccc}
\hline
\hline
  & & & & \\
\multicolumn{1}{l}{Data set} & $\Delta T_0({\rm K})/10^{-4}$ & ${\rm log \, k}$ & $\alpha$ &  $\chi^2$/DOF \\
\hline
 & & & & \\ 
R-FIRAS & $-1.10\pm0.26$ & $-20.49\pm1.90$ & $1.39\pm1.50$ &  2.398\\
 & & & & \\ 
\hline
\end{tabular}
\end{center}
\caption{Results on the parameters of the power law. 
Fits to the FIRAS data with the revised calibration, errors at 95\% CL 
(parabolic approximation errors for ${\rm log \, k}$ and $\alpha$).
We jointly fit $T_0$ and the two parameters, ${\rm log \, k}$ and $\alpha$,
of the astrophysical monopole approximated with a power law.}
\label{tab:fit:0.05}
\end{table*}
\begin{table*}
\begin{center}
\begin{tabular}{lccccc}
\hline
\hline
  & & & & & \\
\multicolumn{1}{l}{Data set} & $\Delta T_0({\rm K})/10^{-4}$ & ${\rm log \, k}_d$ & $\beta$ & $T_d({\rm K})$  & $\chi^2$/DOF \\
\hline
 & & & & &\\ 
R-FIRAS & $-25.45\pm12.89$ & $-18.67\pm0.44$ & $0.91\pm0.15$ & $2.77\pm0.10$  & 1.187\\
 & & & & & \\ 
\hline
\end{tabular}
\end{center}
\caption{Results on the parameters of the dust emission law. 
Fits to the FIRAS data with the revised calibration, errors at 95\% CL 
(parabolic approximation error for $\Delta T_0$ and $\beta$).
We jointly fit $T_0$ and the three parameters, ${\rm log \, k}_d$, $\beta$ and
$T_d$, of the astrophysical monopole approximated with a dust emission law.}
\label{tab:fit:0.05}
\end{table*}

\subsection{Readding the sub-millimetric foreground}

As discussed in the previous subsections, the FIRAS data 
recalibrated according to Battistelli et al. 2000 can not be
fully explained in terms of CMB distortions or of an additional astrophysical 
component. On the other hand, we have carried out this analysis 
by assuming the same subtraction to the (LLSS) FIRAS monopole of the 
astrophysical monopole, $I_{F96}$, derived by Fixsen et~al.~1996 which
is clearly appropriate in the case of a FIRAS calibration 
with an emissivity function very close to unit in the whole LLSS range.

As briefly mentioned in section~2.3, 
the analysis presented in sections~4, 5 and 6 is minimally 
affected by the detailed subtraction of the astrophysical monopole
in the case of an emissivity function very close to unit
at each frequency; this is 
because, after the subtraction of the unperturbed Planckian CMB spectrum,
the astrophysical monopole dominates residual brightness 
at sub-millimetric wavelengths whereas the contribution of the CMB 
spectral distortions dominates at millimetric wavelengths.

Clearly, the calibrator emissivity function considered here calls for 
a relevant millimetric astrophysical foreground or much larger
CMB spectral distortions, that are then relevant also at 
sub-millimetric wavelengths.
We then readd the monopole, $I_{F96}$, derived by Fixsen et~al.~1996 
to the data described in section~7.1 and try 
to fit the data so obtained with a combination of a single astrophysical 
component plus a possibly distorted CMB spectrum.

We start with a single power law plus a pure CMB Planckian spectrum.
The results are reported in Table~10: the $\chi^2$/DOF is clearly high.

We try also a fit in terms of a pure blackbody plus of a dust emission law; 
it gives a very high temperature
$T_d \gsim 10^2$~K (i.e. $h\nu/kT_d \ll 1$)
and $\beta = 0.23\pm0.39$ with $\chi^2$/DOF$\simeq 2.047$,
or, in other words, reduces to the case of a blackbody plus a power law with
$\alpha \simeq \beta + 2$ shown in Table~10, 
(our best fit properly gives $T_d \sim 340$~K and ${\rm log} k_d \sim -22.6$
with quite large errors, because the degeneracy between these two parameters 
in the limit $h\nu/kT_d \ll 1$,
and $\Delta T_0({\rm K})/10^{-4}=-1.62\pm0.45$, in good agreement with the
results of Table~10).
In this subsection we will consider then 
fits with a single power law and 
CMB distorted spectra.

\begin{table*}
\begin{center}
\begin{tabular}{lcccc}
\hline
\hline
  & & & & \\
\multicolumn{1}{l}{Data set} & $\Delta T_0({\rm K})/10^{-4}$ & ${\rm log \, k}$ & $\alpha$ & $\chi^2$/DOF \\
\hline
 & & & & \\ 
R-FIRAS $+$ $I_{F96}$ & $-1.63\pm0.50$ & $-20.23\pm0.48$ & $2.19\pm0.39$ & 2.000\\
 & & & & \\ 
\hline
\end{tabular}
\end{center}
\caption{Results on the parameters of the power law. 
Fits to the FIRAS data with the revised calibration and by readding the 
astrophysical monopole, $I_{F96}$, 
quoted by Fixsen et al. 1996, errors at 95\% CL. 
We jointly fit $T_0$ and the two parameters, ${\rm log \, k}$ and $\alpha$,
of the astrophysical monopole approximated with a power law.}
\label{tab:fit:0.05}
\end{table*}

The results are reported in Tables~11, 12 and 13 respectively 
in the case of an early and a late heating process
and of a combination of them.
Again the data are better explained in terms of 
an early energy injection. On the other hand, 
for a single dissipation process the $\chi^2$/DOF 
remains larger than 1.1 and only with a proper combination
of an early energy injection and a late cooling process
it reduces to a value of about 1.05.

We will discuss in section~7.6 these results from a physical 
point of view.

\begin{table*}
\begin{center}
\begin{tabular}{lccccc}
\hline
\hline
  & & & & & \\
\multicolumn{1}{l}{Data set} & $(\Delta\epsilon/\epsilon_i)/10^{-5}$ & 
$\Delta T_0({\rm K})/10^{-4}$ & ${\rm log \, k}$ & $\alpha$ & $\chi^2$/DOF \\ \\
          & heating at & & & & \\
            & $y_h=5$  & & & & \\
\hline
 & & & & & \\ 
R-FIRAS $+$ $I_{F96}$ & $21.98\pm6.75$ & $-1.15\pm0.33$ & $-21.36\pm0.62$ & $3.09\pm0.50$  & 1.109\\
 & & & & & \\ 
\hline
\end{tabular}
\end{center}
\caption{Results of the fit for the energy injected 
at $y_h=5$ by neglecting late dissipations and free-free distortions 
and for the astrophysical monopole.
Fits to the different data sets with the revised calibration
and by readding the 
astrophysical monopole, $I_{F96}$,
quoted by Fixsen et al. 1996, errors at 95\% CL.
We jointly fit four parameters, $T_0$, the $(\Delta\epsilon/\epsilon_i)$ value referring to 
the early process [$\Ohat_b=0.05$] and 
the two parameters ${\rm log \, k}$ and $\alpha$
of the astrophysical monopole approximated with a power law.}
\label{tab:fit:0.05}
\end{table*}
\begin{table*}
\begin{center}
\begin{tabular}{lccccc}
\hline
\hline
  & & & & & \\
\multicolumn{1}{l}{Data set} & $(\Delta\epsilon/\epsilon_i)/10^{-5}$ & 
$\Delta T_0({\rm K})/10^{-4}$ & ${\rm log \, k}$ & $\alpha$ & $\chi^2$/DOF \\ \\
          & heating at & & & & \\
            & $y_h\ll1$  & & & & \\
\hline
 & & & & & \\ 
R-FIRAS $+$ $I_{F96}$ & $11.94\pm4.74$ & $-0.14\pm0.35$ & $-22.10\pm0.99$ & $3.66\pm0.78$  & 1.604\\
 & & & & & \\ 
\hline
\end{tabular}
\end{center}
\caption{Results on the energy injected 
at $y_h \ll1$ by assuming no early dissipations and a null free-free distortion parameter $y_B$ 
and for the astrophysical monopole.
Fits to the different data sets with the revised calibration
and by readding the 
astrophysical monopole, $I_{F96}$,
quoted by Fixsen et al. 1996, errors at 95\% CL.
We jointly fit four parameters, $T_0$, the $(\Delta\epsilon/\epsilon_i)$ value referring to 
the late process and
the two parameters ${\rm log \, k}$ and $\alpha$
of the astrophysical monopole approximated with a power law.}
\label{tab:fit:0.05}
\end{table*}
\begin{table*}
\begin{center}
\begin{tabular}{lcccccc}
\hline
\hline
  & & & & & & \\
\multicolumn{1}{l}{Data set} & \multicolumn{2}{c}{$(\Delta\epsilon/\epsilon_i)/10^{-5}$} & 
$\Delta T_0({\rm K})/10^{-4}$ & ${\rm log \, k}$ & $\alpha$ & $\chi^2$/DOF \\
\cline{2-3} \\
          & heating at & heating at & & \\
            & $y_h=5$    &  $y_h\ll1$ & & & & \\
\hline
 & & & & & & \\ 
R-FIRAS $+$ $I_{F96}$ & $33.43\pm14.54$ & $-11.31\pm12.79$ & $-2.41\pm0.53$ & $-20.34\pm1.26$ & 
$2.30\pm0.98$  & 1.045\\
 & & & & & & \\ 
\hline
\end{tabular}
\end{center}
\caption{Results on the energy injected 
at $y_h=5$ and at $y_h\ll1$ by assuming a null free-free distortion parameter $y_B$ 
and for the astrophysical monopole.
Fits to the different data sets with the revised calibration
and by readding the 
astrophysical monopole, $I_{F96}$, 
quoted by Fixsen et al. 1996, errors at 95\% CL.
We jointly fit five parameters, $T_0$, 
the two $(\Delta\epsilon/\epsilon_i)$ values referring to 
the early and the late process [$\Ohat_b=0.05$] and 
the two parameters ${\rm log \, k}$ and $\alpha$
of the astrophysical monopole approximated with a power law.}
\end{table*}


\subsection{Joint analysis of CMB spectral distortions and millimetric foreground}

In the previous subsection we have shown that, even allowing for a 
combination of two dissipation processes in the thermal history of the universe,
the recalibrated FIRAS data can not be fully explained
(i.e. the $\chi^2$/DOF remains significantly higher than unit)
when we describe the astrophysical monopole in terms of a single component
particularly important at sub-millimetric wavelengths.
In addition, the results of section~7.3 suggest that a modified blackbody spectrum
at a temperature only just above that of the CMB can play a crucial role 
in the explanation of the recalibrated FIRAS data.

In Table~14 we report then the results of a fit to the 
recalibrated FIRAS data added to the monopole derived by Fixsen et al. 1996
in terms of a pure CMB Planckian spectrum plus an astrophysical
foreground sum of a power law
plus a dust emission law
approximated by a modified blackbody.
As evident, without resorting 
to any kind of CMB spectral distortions,
the addition of a modified blackbody component
significantly improves the fit in a way comparable
to that obtained by including CMB spectral distortions
(the fit quality is intermediate between the case of a single early
process and the case of a proper combination of an early and a late
process in the case of an astrophysical monopole described 
by a simple power law).

\begin{table*}
\begin{center}
\begin{tabular}{lccccccc}
\hline
\hline
  & & & & & & & \\
\multicolumn{1}{l}{Data set} & $\Delta T_0({\rm K})/10^{-4}$ & ${\rm log \, k}$ & $\alpha$ & 
${\rm log \, k}_d$ & $\beta$ & $T_d({\rm K})$ & $\chi^2$/DOF \\
\hline
 & & & & & & & \\ 
R-FIRAS $+$ $I_{F96}$ & $-40.06\pm1.45$ & $-20.56\pm2.93$ & $2.23\pm2.25$ & 
$-18.41\pm0.04$ & $0.79\pm0.11$ & $2.84\pm0.13$  & 1.081\\
 & & & & & & & \\ 
\hline
\end{tabular}
\end{center}
\caption{
Fits to the FIRAS data with the revised calibration
and by readding the 
astrophysical monopole, $I_{F96}$, 
quoted by Fixsen et al. 1996 (parabolic approximation errors at 95\% CL).
We jointly fit six parameters: $T_0$, and the five parameters of 
the astrophysical monopole approximated with 
a power law plus a dust emission law. 
}
\end{table*}

Note that, as found in the fit reported in Table~9, 
by adding a modified blackbody component, the CMB thermodynamic temperature 
$T_0$ recovered by the fit is lower than 2.725~K, 
being $2.725$~K$- T_0 \sim$~some~mK.
The microwave absolute temperature will be then lower at wavelengths of some
millimeters than at millimetric wavelengths, where this modified blackbody component
shows a peak in terms of $\nu I_{\nu}$ (see also Fig.~8), 
whereas at sub-millimetric wavelengths the power law component starts
to dominated the residual brightness after the subtraction of the 
Planckian spectrum at the temperature $T_0$.
Therefore, from a qualitative point of view,  
a modified blackbody component leaves an imprint in the 
microwave spectrum similar to that introduced by an early distortion,
the kind of CMB distortion that better describe the recalibrated FIRAS data.

Therefore, we think meaningless to fit the data jointly in terms of
an additional modified blackbody component and CMB distortions, because 
there is an ``approximate'' degeneration between the 
contributions of these two components to the global monopole. 
Analogously to the approach of section~4, except for the difference
in the assumed astrophysical foreground,
we derive instead the constraints on CMB spectral distortions by keeping
the astrophysical monopole fixed to the description represented 
by the fit results of Table~14.

In Table~15 we report our results in the case of a single
(early, at $y_h \simeq 5$, or late, at $y_h \ll 1$) 
energy dissipation in the thermal history of the universe,
by considering only the frequency range of FIRAS and by adding also 
the recent long wavelength measures of Table~1.
In Fig.~6 we show also the constraints set on a single dissipation process
occurring at arbitrary epochs 
in the thermal history of the universe by the recalibrated FIRAS data 
and compare them with the results based on the calibration 
with a constant emissivity function (i.e. the data $set \, 1)$). As evident,
when the astrophysical monopole is ``properly'' subtracted, the constraints
on the fractional energy exchanges are very similar in the two cases.
The constraints on $(\Delta\epsilon/\epsilon_i)$ 
at very high redshifts are also reported 
in Fig.~3 for a direct comparison with the results based on
the data $set \, 1)$ and $2)$.

In Table~16 we report our results for the case of the joint analysis of 
early ($y_h \simeq 5$) and late ($y_h \ll 1$) 
dissipations, 
by considering only the frequency range of FIRAS and by taking also 
into account the recent long wavelength measures of Table~1.
In Fig.~7 we show also the constraints set on the energy exchanged 
at a given (early or late) epoch by allowing for another 
dissipation process at a different (late or early) epoch
when the recalibrated FIRAS data are combined with 
the recent long wavelength measures.
Again, the comparison with the results based on the calibration 
with a constant emissivity function (i.e. the data $set \, 2)$, see Fig.~3) 
does not show particular differences
in the constraints on $(\Delta\epsilon/\epsilon_i)$. 

Finally, in the case of the FIRAS data calibrated 
with this non constant emissivity function we find a small 
reduction of the $\chi^2$/DOF 
with respect to the case of a constant emissivity function, 
when FIRAS data are considered alone ($\simeq 0.985$ instead of $\simeq 1$)
as well as in combination with long wavelength measures 
($\simeq 1.11$ instead of $\simeq 1.13$).

\begin{table*}
\begin{center}
\begin{tabular}{lcc}
\hline
\hline
  & & \\
\multicolumn{1}{l}{Data set} & \multicolumn{2}{c}{$(\Delta\epsilon/\epsilon_i)/10^{-5}$} \\
\cline{2-3} \\
          & heating at & heating at \\
            & $y_h=5$    &  $y_h\ll1$ \\
\hline
 & & \\ 
R-FIRAS $+$ $I_{F96}$ & $0.15\pm5.35$ & $0.006\pm2.32$ \\
 & & \\ 
$\Delta T_0({\rm K})/10^{-4}$ & $-40.10\pm0.16$ & $-40.10\pm0.17$ \\ 
 & & \\
$\chi^2$/DOF &  0.98514     &   0.98521     \\
 & & \\ 
\hline
 & & \\ 
Table~1 (1985-2000) + [R-FIRAS $+$ $I_{F96}$] & $0.52\pm5.34$ & $0.009\pm2.32$ \\
 & & \\ 
$\Delta T_0({\rm K})/10^{-4}$ & $-40.10\pm0.16$ & $-40.10\pm0.17$ \\ 
 & & \\
$\chi^2$/DOF &  1.09176     &   1.09238    \\
 & & \\ 
\hline
\end{tabular}
\end{center}
\caption{Results on the energy injected at $y_h=5$ and $y_h\ll1$ 
when these two dissipation processes are separately considered.
Fits to the different data sets, errors at 95\% CL.
We jointly fit two parameters: $T_0$ and the values of 
$(\Delta\epsilon/\epsilon_i)$ referring to 
the early or the late process [$\Ohat_b=0.05$].}
\end{table*}
\begin{table*}
\begin{center}
\begin{tabular}{lcccc}
\hline
\hline
  & & & & \\
\multicolumn{1}{l}{Data set} & \multicolumn{2}{c}{$(\Delta\epsilon/\epsilon_i)/10^{-5}$} & $\Delta T_0({\rm K})/10^{-4}$ & $\chi^2$/DOF \\
\cline{2-3} \\
          & heating at & heating at & &\\
            & $y_h=5$    &  $y_h\ll1$ & &\\
\hline
 & & & & \\ 
R-FIRAS $+$ $I_{F96}$ & $0.44\pm9.72$ & $-0.15\pm4.21$ & $-40.11\pm0.30$ & 1.007\\
 & & & & \\ 
Table~1 (1985-2000) + [R-FIRAS $+$ $I_{F96}$] & $1.65\pm9.67$ & $-0.59\pm4.19$ & $-40.14\pm0.29$ & 1.108\\
 & & & & \\ 
\hline
\end{tabular}
\end{center}
\caption{Results on the energy injected at $y_h=5$ and $y_h\ll1$ 
when these two dissipation processes are jointly considered.
Fits to the different data sets, errors at 95\% CL.
We jointly fit three parameters: $T_0$ and the two values of 
$(\Delta\epsilon/\epsilon_i)$ referring to 
the early and late process [$\Ohat_b=0.05$].}
\end{table*}

\begin{figure*}
\epsfig{figure=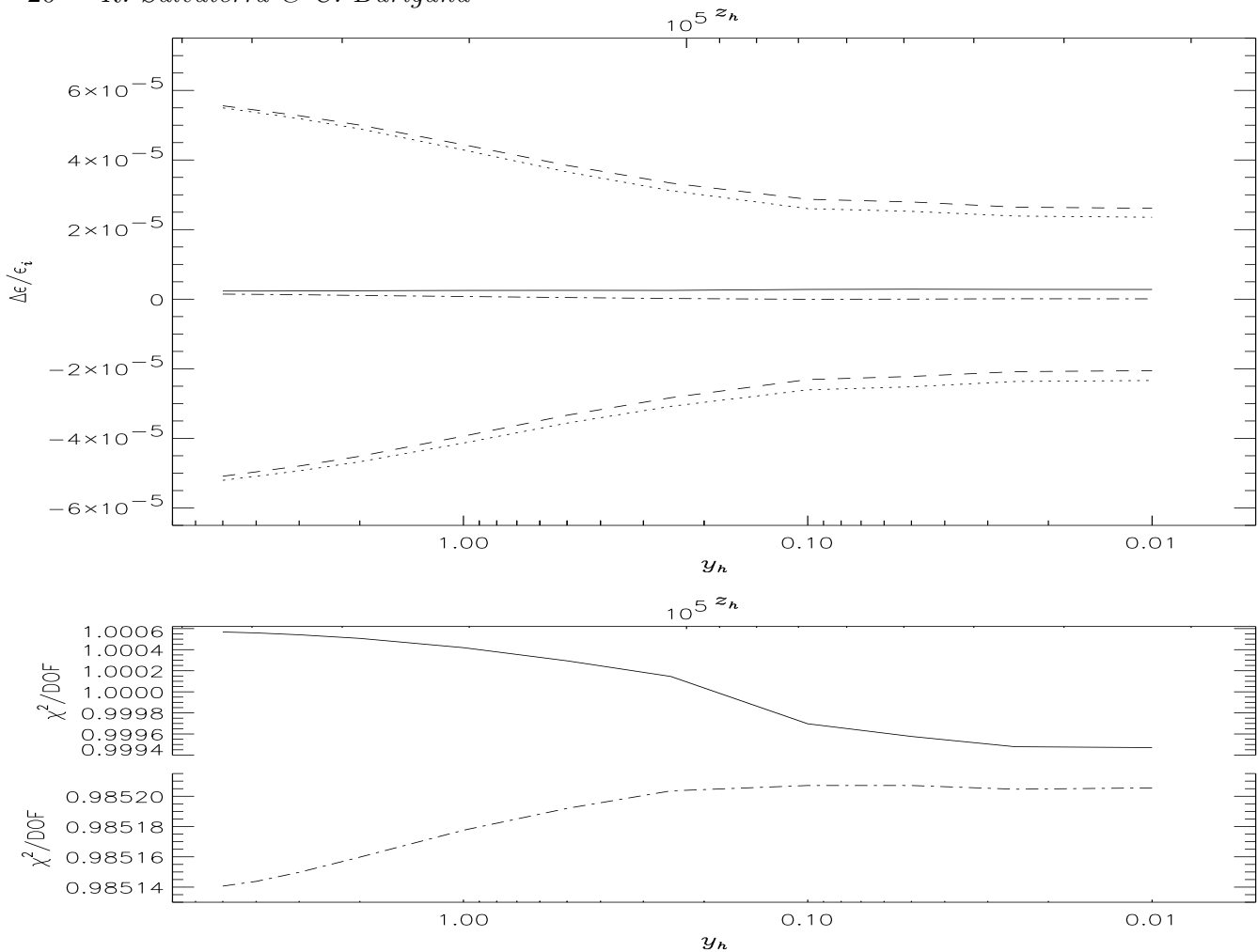,height=13cm,width=17cm}
\caption{The same as in Fig.~1, but with reference
to the exploitation of 
the data $set \, 1)$ (solid lines and dashed lines)
and of the FIRAS data alone but calibrated according to 
Battistelli et al.~(2000) by adding the ``experimental'' astrophysical 
monopole, $I_{F96}$, 
derived by Fixsen et al. (1996) and subtracting the ``theoretical'' 
astrophysical monopole of the best fit obtained assuming a power law plus a 
dust emission law as in Table ~14 (dot-dashed lines and dotted lines)
[$\Ohat_b = 0.05$].
}
\end{figure*}
\begin{figure*}
\epsfig{figure=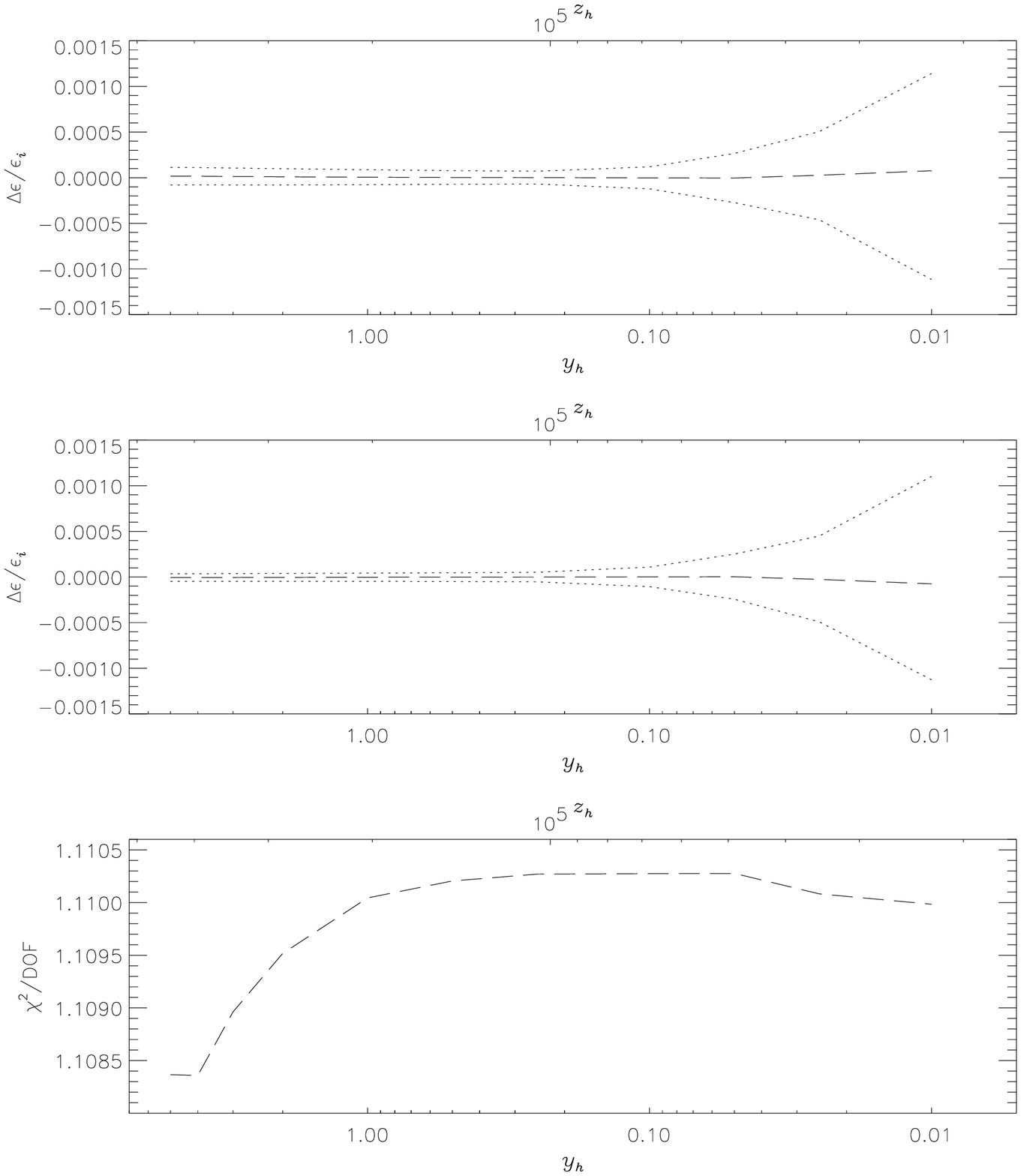,height=13cm,width=17cm}
\caption{The same as in Fig.~3, but with reference
to the exploitation of 
the FIRAS data alone but calibrated according to 
Battistelli et al.~(2000) by adding the ``experimental'' astrophysical 
monopole, $I_{F96}$,  
derived by Fixsen et al. (1996) and subtracting the ``theoretical'' 
astrophysical monopole of the best fit obtained assuming a power law plus a 
dust emission law (see Table~14)
jointed to the recent ground and balloon measures 
[$\Ohat_b = 0.05$].
}
\end{figure*}

\subsection{Cosmological and astrophysical implications}

The explanation of the FIRAS data recalibrated according to
Battistelli et al. (2000) in terms of pure CMB spectral distortions
seems to be difficult. From the point of view of the data analysis,
the better solution in this scheme involves a small modification
of the astrophysical monopole with respect to that given 
by Fixsen et al. 1996, $I_{F96}$, (clearly compatible with the 
current models for the sub-millimetric extragalactic foreground
generated by distant galaxies)
and a relevant early energy injection,
$\Delta \epsilon/\epsilon \sim 3 \times 10^{-4}$,
which effect on the spectrum
has to be partially modified by a late cooling process
associated to an energy exchange smaller by a factor $\sim 3$,
$\Delta \epsilon/\epsilon \sim 10^{-4}$,
to give a $\chi^2$/DOF quite close to unit (see Tables~11 and 13).
We have also tested that an energy injection at intermediate 
values of $y_h$ does not improve but worses the fit quality with respect to
the results of Tables~11 and 13.
A vacuum decay with a radiative channel can not explain
this result because it predicts a late positive energy dissipation
much larger than the early one (see, e.g., Freese et al. 1987, Weinberg 1988, 
Bartlett \& Silk 1989, Overduin et al. 1993).
The damping of adiabatic perturbations (Sunyaev \& Zeldovich 1970) may generate 
a relevant early distortion (Daly 1991, Barrow \& Coles 1991, 
Burigana 1993, Hu et al. 1994), but, unfortunately, 
only for cosmological parameters excluded by the recent
CMB anisotropy experiments (see, e.g., Netterfield et~al. 2002, Stompor et~al. 2001, 
Pryke et~al. 2002, and references therein).
Also the damping of isocurvature perturbations 
seems to be not able to generate such kind of distortion,
because the high density contrast predicted at high redshifts 
is expected to significantly increase the Bremsstrahlung efficiency 
in erasing early distortions, 
so that the generation of a late distortion larger than 
the early one is favourite (Daly 1991, Burigana 1993).
Radiative decays of massive particles (Silk \& Stebbins 1983)
at early times, with appropriate number density, mass, lifetime 
and branching ratio may produce such kind of early distortion.
On the other hand, a proper balance between the parameters 
of the early decay and of a completely different cooling
process at late epochs or a delicate fine tuning 
of the decay parameters at intermediate epochs, 
able to accurately reproduce the required 
spectral shape (see, e.g., Danese \& Burigana 1993)
without affecting the evolution of the red giant branch 
(Raffelt et~al. 1989), is required for a full
explanation of the data, otherwise 
the $\chi^2$/DOF is larger than unit of about the 10\%.
Of course these possibilities can not be excluded, 
but they seem quite unrealistic or, at least, quite weak.

Measurements at $\lambda \approx 5.64 \Ohat_b^{-2/3}$~cm
may in principle play a relevant role for the understanding
of early distortions, being there maximum the amplitude 
of the BE-like distortions, 
$(\Delta T/T)_m \simeq 5.82 \mu (y_h \simeq 5) \Ohat_b^{-2/3}$
(Burigana et al. 1991a).
Unfortunately, the error 
bars~\footnote{According to the considered frequency and 
observational method,
the major contributions to the experimental uncertainty 
of long wavelength spectrum measurements derive 
from instrumental effects and/or atmospheric and foreground contamination
(see, e.g., Sironi et al. 1990, 1991, Bensadoun et al. 1993, 
Bersanelli et al. 1994, Salvaterra \& Burigana 2000, and references therein).}
of the current measures at $\lambda \gsim 20$~cm
are much larger (by about one order of magnitude, see Table~1) than the maximum 
distortion of about 50--100~mK compatible with the upper limit 
(at 95\% CL) on $\mu (y_h \simeq 5)$ ($\simeq 1.4 \Delta \epsilon/\epsilon_i$)
derived here in the case of the calibration of FIRAS data by Battistelli et al. (2000)
for $\Ohat_b$ compatible with recent CMB anisotropy experiments 
and the big-bang nucleosynthesis.
A great improvement of long wavelength spectrum measurements 
is therefore necessary for consistency tests of FIRAS calibration. 

If we allow for a further component to the astrophysical foreground,
relevant at millimetric wavelengths, 
quite well approximated 
by a dust emission law with modified blackbody parameters 
close to those of Table~14 (to be considered only as 
a simple set of observational/phenomenological 
parameters), the recalibrated FIRAS data 
can be properly explained by including 
CMB spectral distortions generated by dissipation processes  
with energy exchanges
($\Delta \epsilon/\epsilon \approx 10^{-5}$, compatible
with null values) consistent with current 
cosmological scenarios, both at early and late epochs,
without requiring a proper balance between 
the energy possibly injected at different cosmic times to 
obtain a $\chi^2$/DOF extremely close to unit.
We consider then the astrophysical implications of this scheme.

In Fig.~8 we compare the brightness 
of the dust emission component, $I_{Dbf}$, 
represented by the best fit reported in Table~14 with 
the sub-millimetric foreground as derived by Fixsen et al. 1996 
and by Fixsen et al. 1998,
the predictions based on the models by Toffolatti et al. 1998 and 
by Guiderdoni et al. 1998, the brightness of the Galaxy at Pole as found
by Fixsen et al. 1996 
(extended in the figure up to sub-millimetric frequencies) 
and the power law component represented by the best fit 
reported in Table~14.

\begin{figure*}
\epsfig{figure=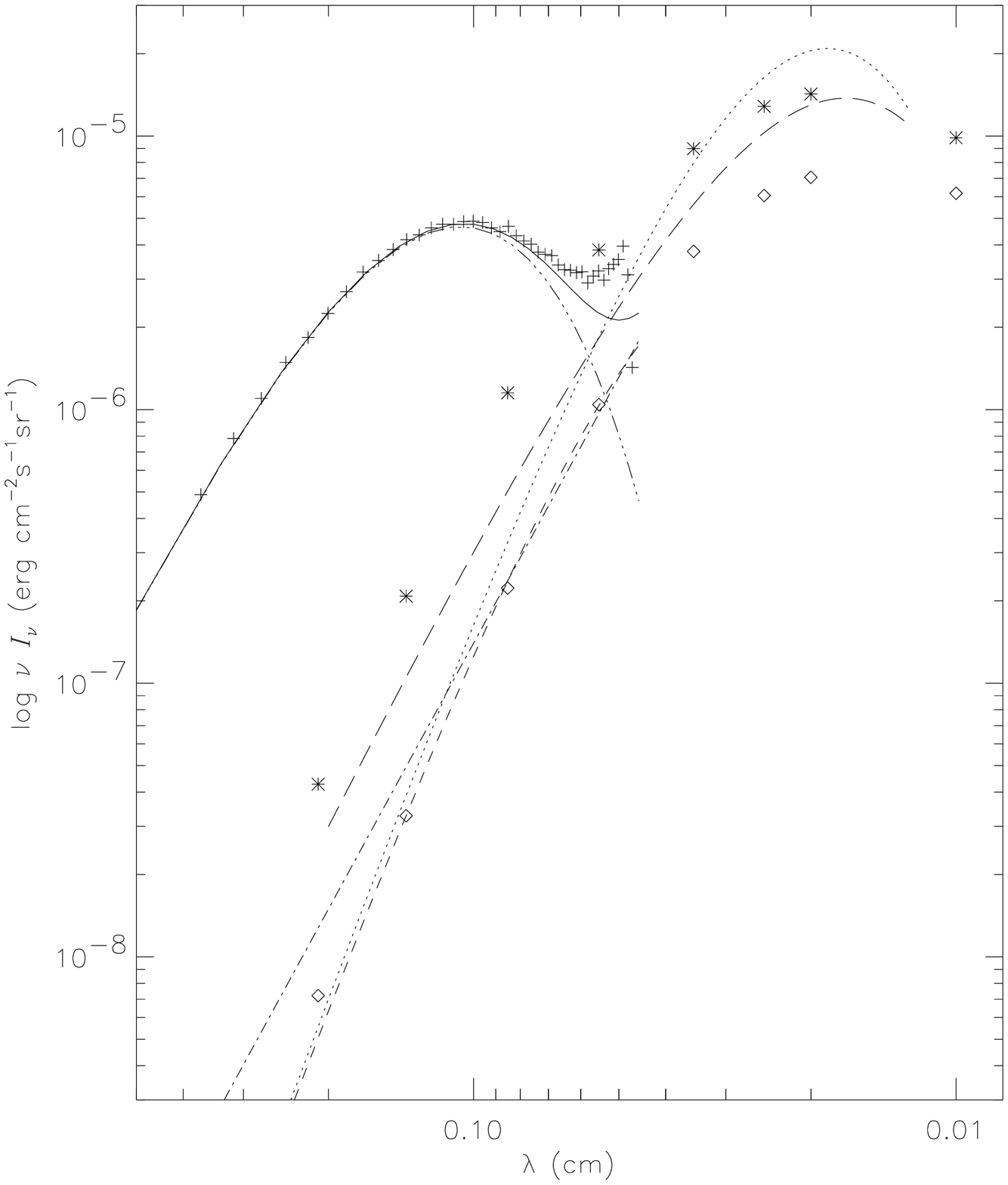,height=10cm,width=17cm}
\caption{Comparison between the brightness 
of the dust emission component, $I_{Dbf}$, 
of the best fit reported in Table~14 (three dots/dash) and
the sub-millimetric foreground as derived by Fixsen et al. 1996 (dashed line)
and by Fixsen et al. 1998 (long dashes),
the predictions based on the models of Toffolatti et al. 1998 (diamonds) and 
of Guiderdoni et al. 1998 (asteriscs), the brightness of the Galaxy at Pole as found
by Fixsen et al. 1996 (extended in the figure up to sub-millimetric frequencies,
dotted line) 
and with the power law component, $I_{PLbf}$,  represented by the best fit 
reported in Table~14 (dot-dashed line).
We report also the sum of $I_{Dbf}$ and  $I_{PLbf}$ (solid line)
compared to the LLSS FIRAS data of Fixsen et al. 1996, including the monopole $I_{F96}$,
but calibrated according to Battistelli et al. 2000 after the subtraction of a blackbody
at the best fit temperature $T_0 \simeq 2.721$ reported in Table~14 (crosses).
The $1\sigma$ uncertainty of these data is not reported being less
than $\sim 10^{-7}$ ${\rm erg} \, {\rm cm}^{-2} {\rm sec}^{-1} {\rm sr}^{-1}$ 
at the wavelengths $\lambda \gsim 700 \mu$m relevant for the evaluation
of the considered millimetric foreground. As evident, the shape and level of 
$I_{Dbf}$ at $\lambda \gsim 700 \mu$m does not critically depend on the 
details of the subtraction of the sub-millimetric contribution from distant galaxies. 
}
\end{figure*}

The comparison of the brightness integrated over the frequencies 
of the dust emission component $I_{Dbf}$ 
with the CMB integrated brightness gives
$\int I_{Dbf} d\nu / \int B_{\nu}(T_0) d\nu \sim 1/170 \sim 4 \Delta T_0 /T_0$ 
(where $\Delta T_0 \simeq $~4mK, see Table~14); this value, much larger than
the upper limits on $\Delta \epsilon/\epsilon_i$ reported 
in Tables~$5 \div 7$ and $11 \div 13$,
is simply due to the lower best fit value of $T_0$ found in this case.
Analogously, the brightness integrated over the frequencies 
of the dust emission component $I_{Dbf}$ 
is about 1/4 of the integrated brightness 
of the Galaxy at Pole component found by Fixsen et al. 1996 
(extrapolated up to sub-millimetric wavelengths),
$\int I_{Dbf} d\nu / \int I_{GP} d\nu \sim 1/4$.
By comparing $I_{Dbf}$ with the 
the extragalactic sub-millimetric 
foreground generated by the distant galaxies, as derived 
by Fixsen et al. 1998, we find
$\int I_{Dbf} d\nu / \int I_{F98} d\nu \sim 1/3$. 
So, there are no particular problems from the energetic point of view
for an explanation of this millimetric component in terms of emission from
cold dust 
in an extended halo around our Galaxy or 
around distant galaxies; in the latter case 
the high isotropy level of this component 
at angular scales larger than few degrees is simply explained.
Unfortunately, a direct detection of millimetric emission
quite far from the central regions of distant dusty galaxies seems 
very difficult with current millimetric telescopes.

Assuming that a fraction $f$ of the cold dust producing
this millimetric foreground is located 
around distant dusty galaxies,
an approximate estimate of the corresponding Poisson rms fluctuation
level, $\sigma_{Dbf}$, can be simply derived by rescaling 
the usual rms fluctuation, $\sigma_{dg}$, of dusty galaxies.
For differential number counts 
approximately described by a power-law form,
$dN(S)/dS \simeq g S^{-(\gamma+1)}$, 
the rms fluctuation is given by 
$\simeq q^{(2-\gamma)/\gamma} (g\omega_e)^{1/\gamma} 
/(2-\gamma)^{1/\gamma}$, where $\omega_e$ is the antenna beam 
pattern effective solid angle 
and sources above the chosen $q-\sigma$ clipping detection limit
(being $\sigma$ the rms confusion noise from all contributions 
and $q \sim 2.5 \div 5$) are removed (De Zotti et al. 1996).
The ratio between the level of the contribution to this
millimetric foreground associated to halos around distant dusty galaxies
and the sub-millimetric foreground extrapolated at millimetric wavelengths 
is given by $r f$, where $r$
ranges from $\sim 20$ to $100$ for $\nu$ from $\sim 300$ to 100~GHz 
(see Fig.~8).
Assuming the same typical ratios, $r f$, for the 
corresponding source intrinsic components,
we have that the value of $g$ relevant for the millimetric number counts,
 $g_{Dbf}$, is related to that relevant for the sub-millimetric number counts,
 $g_{dg}$, by $ g_{Dbf} \sim (r f)^{(\gamma+1)} g_{dg}$, and then
$\sigma_{Dbf} \sim (r f)^{(\gamma+1)/\gamma} \sigma_{dg}$.
Assuming $\gamma = 1.25 \div 1.5$ and the current estimates
of the combined rms fluctuation from dusty galaxies and radiogalaxies, 
$\sigma_{ex}$,  
at these frequencies (Toffolatti et al. 1999), we find
$\sigma_{Dbf} \sim (20 \div 15) \sigma_{ex}$
(or $\sigma_{Dbf} \sim (4.5 \div 3) \sigma_{ex}$)
even for $f \sim 0.25$ (or $\sim 0.1$). 
A significant excess in the angular 
power spectrum at high multipoles 
is then unavoidable in this scheme, 
unless galactic outflows and/or interactions 
efficiently redistribute the cold dust more uniformly 
in the intergalactic medium.
It should be 
observable with the current space anisotropy experiment 
MAP\footnote{http://map.gsfc.nasa.gov/} by NASA, 
and, with a wide frequency coverage, with the {\sc Planck}
satellite\footnote{http://astro.estec.esa.nl/Planck} by ESA
even for quite small values of $f$ ($f \sim 0.07$).

On the other hand, the implications for the dust mass 
involved in this scenario is critical.
The value of ${\rm k}_d$ derived from the fit (see Table~14)
is about 1500 times larger than the analogous parameter 
corresponding to the monopole, $I_{F98}$, derived by 
Fixsen et al. 1998. Therefore, the involved mass of cold dust
around distant galaxies (or its intrinsic emissivity)
should be orders of magnitude larger than 
that producing the sub-millimetric foreground.
Of course, an analogous mass problem holds for a cold dust halo 
around the Galaxy.
This problem can be overcome by considering cold dust at 
low redshifts ($z \approx 0.1$), but in this case 
the anisotropy problem described above seems even more 
critical, since an efficient dust redistribution at low redshifts
is unrealistic; in addition, we should find observational 
evidences at low redshifts.

\section{Conclusions}

In this work we have compared the absolute temperature data 
of the CMB spectrum with models for CMB spectra distorted by a single or two 
heating processes at different cosmic times. 

We have computed the 
limits on the amount of the energy injected in the radiation field
for the whole range of cosmic epochs, expressed here in terms of the 
dimensionless time variable $y_h$.
These upper and lower limits on  
$\Delta\epsilon/\epsilon_i$ are mainly provided by the precise measures of the 
FIRAS instrument aboard the COBE satellite.
The addition of the data obtained from
ground and balloon experiments at longer wavelengths does not 
alter significantly the results based on the FIRAS data alone, because of
the large error bars of the measures at $\lambda \gsim 1$~cm.
We analyzed also the impact of the FIRAS
calibration on the determination of $\Delta\epsilon/\epsilon_i$ when the
FIRAS data are used together with the ground and balloon measures:
the uncertainty of 2~mK at 95\% CL in the
FIRAS calibration (Mather et al. 1999) does not affect significantly the limits
on $\Delta\epsilon/\epsilon_i$. 
From the $\chi^2$ analysis, the lower limit at 2.723~K results weakly
favourite.
This is due to the well known disagreement
between the absolute temperature of the FIRAS data and of the mean temperature of 
the data at $\lambda>1$ cm.

We considered different values of the baryon density.
For the same $y_h$, the value of $\Omega_b$ does not significantly influence 
the upper and lower limits on the amount of the injected energy
(of course, $z_h(y_h)$ decreases with $\Omega_b$).

As in the case of a single heating, we exploit the CMB spectrum data 
under the hypothesis of two heating processes 
occurred at different epochs, 
the former at any $y_h$ in the range $5 \geq y_h \geq 0.01$ 
(but only for $y_h \gsim 0.1$ this analysis results to be meaningful)
and the 
latter at $y_h \ll 1$. 
The limits on $\Delta\epsilon/\epsilon_i$ are relaxed by a factor $\sim 2$ 
both for the earlier and the later process with respect to the case in which 
a single energy injection in the thermal history of the universe is considered.

Also in this case, we analyzed the impact of the FIRAS calibration and the role
of the baryon density. The results are very similar to those obtained 
for the case of single heating.

In general, the constraints on  $\Delta\epsilon/\epsilon_i$
are weaker for early processes
($5 \gsim y_h \gsim 1$) than for relatively late processes
($y_h \lsim 0.1$), because of the sub-centimetric wavelength coverage
of FIRAS data, relatively more sensitive to Comptonization
than to Bose-Einstein like distortions.

We evaluate also the limits on $\Delta\epsilon/\epsilon_i$ 
for energy injections occurring during the kinetic equilibrium period
(i.e. at $z \gsim z_1$). By allowing for a further late dissipation process,
the constraints on $\Delta\epsilon/\epsilon_i$ at $z \gsim z_1$ 
return to be relaxed with respect to the case of a 
single injection at high $z$, 
particularly at $z_1 \lsim z \lsim z_{therm}/2$
(up to a factor $\sim 2$).

In conclusion, the available data permit to set very stringent 
constraints on the energy injected in the radiation field
at different cosmic times, mainly set
by the precise measures of FIRAS.

On the other hand, 
measures at $\lambda \gsim 1$~cm, where significant improvements
are needed, play a crucial role to probe free-free distortions
and to better constrain the thermal history of the universe
at late epochs.

We carefully considered also the implications of the 
FIRAS calibration as revised by Battistelli et al. 2000.
From a widely conservative point of view (i.e. allowing for the
whole set of calibrator emissivity curves reported by
the authors), it only implies a significant relaxation 
of the constraints on the Planckian shape of the CMB spectrum.
On the other hand, the favourite calibrator emissivity law
proposed by the authors, when used to recalibrate the FIRAS data, 
implies significant deviations from a Planckian spectrum.
If interpreted in terms of CMB spectral distortions, the full 
data set can be explained only by assuming 
a proper balance between the energy exchanges at two 
completely different cosmic times 
or a delicate fine tuning 
of the parameters of a dissipation process at intermediate 
epochs, possibly in form radiative decays of massive particles. 
The interpretation in terms of a relevant millimetric 
foreground, quite well described by a modified blackbody law
as in the case of emission from cold dust,
does not present this problem, the constraints on energy 
exchanges in the primeval plasma being close to those
derived above by assuming a constant emissivity of the calibrator.
On the other hand, the too large mass of dust required in this scenario
and/or the significant increase of the fluctuations at sub-degree
angular scales represent again a very difficult problem.
While a careful control of the calibration of CMB spectrum
observations at this high level of sensitivity is essential 
to establish limits on (or measure) $\Delta\epsilon/\epsilon_i$
with an accuracy of $\sim 10^{-5} \div 10^{-4}$, 
our analysis indicates that it is very difficult to explain
a non constant FIRAS calibrator emissivity law 
with a spectral dependence close to that considered here.


Future precise measurements at longer wavelengths,
particularly significant for early dissipation processes,
as well as current and future CMB anisotropy space missions
will provide independent, direct or indirect, cross checks.


\section*{Acknowledgements}

It is a pleasure to thank M.~Bersanelli, D.J.~Fixsen, N.~Mandolesi, 
C.~Macculi and G.~Palumbo 
for useful discussions on CMB spectrum observations.
We warmly thank L.~Danese, G.~De~Zotti and L.~Toffolatti
for constructive and stimulating comments and 
the fruitful long-term collaboration on CMB spectral distortions 
and astrophysical foregrounds. We wish to thank the referee
for valuable comments.

\end{document}